\documentclass[fleqn,usenatbib]{mnras}

\usepackage{newtxtext,newtxmath}

\usepackage[T1]{fontenc}
\usepackage{ae,aecompl}

\usepackage{graphicx}	
\usepackage{amsmath}	
\usepackage{placeins}
\usepackage{float} 
\usepackage{longtable}

\newcommand{\source}{GX~339$-$4}

\newcommand{\Swift}{{\it Swift}}

\newcommand{\ergscm}{erg\,s$^{-1}$\,cm$^{-2}$}
\newcommand{\swift}{\textit{Swift}}

\newcommand{\lEdd}{$L_{\rm Edd}$}

\newcommand{\nh}{$N_{\rm H}$}

\graphicspath{{./}{figures/}}

\title[Hard-only outbursts from \source]{Disk--jet coupling changes as a possible indicator for outbursts from \source\,remaining within the X-ray hard state}

\author[S. E. M. de Haas et al.]{S. E. M. de Haas,$^{1,2}$\thanks{E-mail: sebastiaan.dehaas@student.uva.nl}
T. D. Russell,$^{3,1}$
N. Degenaar,$^{1}$
S. Markoff,$^{1,2}$
A. J. Tetarenko,$^{4}$
\newauthor
B. E. Tetarenko,$^{5}$
J. van den Eijnden,$^{1,6}$
J. C. A. Miller-Jones,$^{7}$
A. S. Parikh,$^{1}$
R. M. Plotkin,$^{8}$
\newauthor 
and G. R. Sivakoff$^9$\\
$^{1}$Anton Pannekoek Institute for Astronomy, University of Amsterdam, Science Park 904, 1098 XH Amsterdam, The Netherlands\\
$^{2}$Gravitation Astroparticle Physics Amsterdam (GRAPPA) Institute, University of Amsterdam, Science Park 904, 1098 XH Amsterdam, The Netherlands\\
$^{3}$INAF, Istituto di Astrofisica Spaziale e Fisica Cosmica, Via U. La Malfa 153, I-90146 Palermo, Italy\\
$^{4}$East Asian Observatory, 660 N. A'oh\={o}k\={u} Place, University Park, Hilo, Hawaii 96720, USA\\
$^{5}$Department of Astronomy, University of Michigan, 1085 South University Avenue, Ann Arbor, MI 48109, USA\\
$^{6}$Astrophysics, Department of Physics, University of Oxford, Denys Wilkinson Building, Keble Road, Oxford OX1 3RH, UK\\
$^{7}$International Centre for Radio Astronomy Research - Curtin University, GPO Box U1987, Perth, WA 6845, Australia\\
$^{8}$Department of Physics, University of Nevada, Reno, Nevada, 89557, USA\\
$^{9}$Department of Physics, University of Alberta, CCIS 4-181, Edmonton, AB T6G 2E1, Canada\\
}

\date{Accepted 4 December 2020.}

\pubyear{2020}

\begin{document}
\label{firstpage}
\pagerange{\pageref{firstpage}--\pageref{lastpage}}
\maketitle

\begin{abstract}
\noindent

We present quasi-simultaneous radio, (sub-)millimetre, and X-ray observations of the Galactic black hole X-ray binary \source, taken during its 2017--2018 outburst, where the source remained in the hard X-ray spectral state. During this outburst, \source\ showed no atypical X-ray behaviour that may act as a indicator for an outburst remaining within the hard state. However, quasi-simultaneous radio and X-ray observations showed a flatter than expected coupling between the radio and X-ray luminosities (with a best fit relation of $L_{\rm radio} \propto L_{\rm X}^{0.39 \pm 0.06}$), when compared to successful outbursts from this system ($L_{\rm radio} \propto L_{\rm X}^{0.62 \pm 0.02}$). While our 2017--2018 outburst data only span a limited radio and X-ray luminosity range ($\sim$1 order of magnitude in both, where more than 2-orders of magnitude in $L_{\rm X}$ is desired), including data from other hard-only outbursts from \source\ extends the luminosity range to $\sim$1.2 and $\sim$2.8 orders of magnitude, respectively, and also results in a flatter correlation (where $L_{\rm radio} \propto L_{\rm X}^{0.46 \pm 0.04}$). This result is suggestive that for \source\ a flatter radio -- X-ray correlation, implying a more inefficient coupling between the jet and accretion flow, could act as an indicator for a hard-only outburst. However, further monitoring of both successful and hard-only outbursts over larger luminosity ranges with strictly simultaneous radio and X-ray observations is required from different, single sources, to explore if this applies generally to the population of black hole X-ray binaries, or even \source\ at higher hard-state luminosities.

\end{abstract}


\section{Introduction}

Black hole (BH) low-mass X-ray binaries (LMXBs) are systems where material from a low-mass ($< 3\,M_{\odot}$) donor star is transferred to a stellar-mass black hole via Roche lobe overflow. The infalling matter forms a differentially rotating accretion disk \citep[e.g.,][]{1973A&A....29..179P} and material is partially accreted onto the BH, while the remaining fraction may be ejected from the system via outflows in the form of relativistic jets \citep[e.g.,][]{1992Natur.358..215M,2001MNRAS.322...31F} or disk winds \citep[e.g.,][]{2012MNRAS.422L..11P,2016AN....337..368D}. 
LMXBs are ideal objects for studying accretion and jet physics as they evolve on observable (month-year) time scales, providing a time-resolved view of their evolution through their outburst duty cycles (which, are phases of increased accretion onto the BH) and the connection between the inflow and outflow as they evolve \citep[e.g.,][]{2010LNP...794..115F}. These objects also allow us to observe their interaction with their surroundings (e.g.,\
\citealt{2002ApJ...573L..35C};
\citealt{2004ApJ...617.1272C};
\citealt{2005Natur.436..819G};
\citealt{2011A&A...528A.149M};
\citealt{2012MNRAS.423.1641J};
\citealt{2017MNRAS.468.2788R};
\citealt{2018MNRAS.475..448T};
\citealt{2019ApJ...883..198R};
\citealt{2020NatAs...4..697B};
\citealt{2020MNRAS.497.3504T}).

BH LMXBs emit across the electromagnetic spectrum, where the near-Infrared (IR) and optical to X-ray radiation typically originates from the in-flowing material (often with some contribution from the companion star or disk winds), while the radio, millimetre (mm) and far-infrared emission is generally radiated by the jet \citep[e.g.,][]{2001MNRAS.322...31F,2013MNRAS.429..815R,2014MNRAS.439.1390R,2015ApJ...805...30T}. These objects spend the majority of their lifetimes in a low-luminosity quiescent state (with X-ray luminosities of $\lesssim 10^{-5}$ $L_{\rm Edd}$; \citealt{plotkin2013}). However, they may go through episodic phases of outburst that typically last months to $\sim$a year, during which their luminosities can increase by orders of magnitude due to an increase in the mass-accretion rate onto the BH
\citep[see, e.g.,][]{2006csxs.book..157M,watchdog,2016A&A...587A..61C}.

During quiescence and the early rise phase of an outburst, the system is typically in the hard accretion state. In this state, the observed X-ray spectrum is dominated by a power-law component, thought to arise from inverse Compton emission originating in a geometrically-thick, optically-thin, radiatively inefficient accretion flow \citep[RIAF; e.g.,][]{1995ApJ...452..710N,1997ApJ...489..865E}, or possibly the base of the jet \citep[e.g.,][]{2005ApJ...635.1203M,2009MNRAS.398.1638M,2019MNRAS.485.3696C,2019MNRAS.482.4798L}. The quiescent and hard state are associated with a steady, partially self-absorbed compact jet \citep[e.g.,][]{fender2004} exhibiting a flat-to-inverted radio to (sub-)mm spectrum (e.g.,\ \citealt{2000ApJ...543..373D,2000A&A...359..251C,2001MNRAS.322...31F, 2014MNRAS.439.1390R}, \citealt{2015MNRAS.450.1745R,2015ApJ...805...30T}, \citealt{2017ApJ...848...92P,2018ApJ...852....4D}), which can extend up to and beyond the IR band \citep[e.g.,][]{2013MNRAS.429..815R}. For such a spectrum, the radio flux density $S_{\nu}$ is proportional to the frequency $\nu$ such that $S_{\nu} \propto \nu^{\alpha}$, where the spectral index $\alpha \gtrsim 0$ \citep[e.g.,][]{fender2006}. 

As a typical outburst progresses, it is generally thought that material that has built in the outer disk up over time reaches a critical point, following which, material can then flow in towards the BH, causing the source to source brighten \citep[e.g.,][]{2001NewAR..45..449L}. During this phase, the X-ray spectrum progressively softens as the X-ray emission becomes increasingly dominated by a multi-temperature blackbody component from the hot inner regions of an optically-thick, geometrically-thin accretion disk. As a consequence, the system may enter and transit through the hard intermediate and soft intermediate X-ray states (HIMS and SIMS, respectively; e.g.,  \citealt{2006ARA&A..44...49R,2010LNP...794...53B}), entering the full soft state as the disk completely dominates the observed X-ray emission. At some point during this progression, the compact jet emission is strongly quenched (by $>$3.5\,orders of magnitude; \citealt{2019ApJ...883..198R}) and discrete ejecta may be launched in a transient jet \citep[e.g.,][]{1994Natur.371...46M,1995Natur.375..464H,fender2006}. These transient jet ejecta are thought to be composed of steep spectrum ($\alpha \approx -0.7$) discrete, synchrotron emitting plasma that are propagating away from the BH at relativistic speeds \citep[e.g.,][]{1995Natur.374..141T,2009MNRAS.396.1370F,2017MNRAS.469.3141T} and typically manifest as rapid flaring activity in time-resolved light curves. 

Following the spectral transition, the luminosity drops as the mass accretion rate onto the compact object reduces and the outburst fades. As the source decays it transits back through the intermediate states to the hard state in a reverse transition \citep[e.g.,][]{1996ApJ...457..821N,1997ApJ...489..234H}. 
During this progression, the compact jet switches back on progressively
(over a period of a few weeks; e.g.,\ \citealt{2012MNRAS.421..468M,2013ApJ...779...95K,2013MNRAS.431L.107C,2014MNRAS.439.1390R}).

However, not all outbursts follow this typical ``successful'' progression. Instead, outbursts can exhibit multiple peaks, re-brightenings, or glitches \citep[e.g.,][]{1997ApJ...491..312C}, during which the connection to the jet is not well understood \citep[e.g.,][]{2019ApJ...878L..28P}. Some outbursts fail to enter the soft state, such that the source remains in the hard X-ray spectral state \citep[e.g.,][]{1994ApJ...425L..17H,2000ApJ...539L..37H,2001MNRAS.323..517B,2004NewA....9..249B,2010MNRAS.404..908B,2002A&A...390..199B,2004AstL...30..669A,2005ApJ...625..923S,2011MNRAS.415.2373S,2013A&A...557A..45C} or only progresses as far as the intermediate states during the outburst \citep[e.g.,][]{2002A&A...394..553I,2002ApJ...564..974W,2009MNRAS.398.1194C,2012A&A...537L...7F,2012ApJ...751...34R,2013MNRAS.429.1244S,2013MNRAS.431.2285Z,2014MNRAS.437.3265C}. These outbursts are generally referred to as ``hard-only'' outbursts and it is not well understood why some outbursts do not progress to a soft state. One possible explanation is that hard-only outbursts are a result of less disk material being able to flow into the inner regions of the accretion disk, such that a full outburst cannot be sustained. As would be implied by such a scenario, hard-only outbursts are typically fainter on-average than successful outbursts \citep{watchdog}, indicating lower peak accretion rates. Additionally, hard-only outbursts are shorter in duration, lasting on average $\sim$247 days compared to $\sim$391 days for successful outbursts \citep{watchdog}.

In the hard states of BH LMXB outbursts, there is an observed correlation between the radio luminosity ($L_{\rm r}$) and X-ray luminosity ($L_{\rm X}$). This correlation holds over a few orders of magnitude in luminosity, and can be used to investigate the disk--jet coupling \citep[e.g.,][]{1998NewAR..42..601H,2000A&A...359..251C,2003A&A...400.1007C,corbel,2003MNRAS.344...60G,gallotracks,2018MNRAS.478L.132G}. The observed correlation is non-linear and typically divided by two separate tracks, the upper `standard' track and a lower, shallower one, although the statistical significance of the need for two different tracks is debated
\citep{2014MNRAS.445..290G,2018MNRAS.478L.132G}. It has also been suggested that the difference between the two slopes may be related to the observed jet emission \citep[e.g.,][]{2009ApJ...703L..63C,2011MNRAS.414..677C,2014A&A...562A.142M,2018MNRAS.473.4122E}. According to \cite{gallotracks}, 
for the full sample of standard track BH LMXBs, the radio luminosity
correlates with X-ray luminosity as $L_{\rm r} \propto L_{\rm X}^{0.63 \pm 0.03}$. Using only data from \source\ over multiple outbursts (including both successful and hard-only outbursts), \citet{corbel} reported a relation of $L_{\rm r} \propto L_{\rm X}^{0.62 \pm 0.01}$. However, \citet{corbel} show that over limited X-ray luminosity ranges ($<$2 orders of magnitude) individual outbursts can show significant deviation from the source's typical correlation.

In this paper we present X-ray, radio and (sub-)mm monitoring of the Galactic BH LMXB \source\ during its 2017--2018 hard-only outburst. In Section~\ref{sec:obs} we describe the observational setup and data used. Section~\ref{sec:results} presents the results from the radio, (sub-)mm, and X-ray monitoring. In Section~\ref{sec:discussion}, we discuss and compare the results: exploring the X-ray spectral evolution, in particular the source hardness and X-ray photon index ($\Gamma$), as well as the behaviour of the radio jet. We also attempt to physically explain the observed evolution of this outburst. Finally, we compare our results to other outbursts of \source, both hard-only and successful, searching for any X-ray or radio signatures that may act as indicators for an outburst only remaining in the hard or intermediate states and not progressing into the soft state. Conclusions are provided in Section~\ref{sec:conclusion}. All radio, (sub-)mm and X-ray results are presented throughout the paper, and in the Appendices.

\subsection{\source}
\label{gx3394intro}
\source\ is a LMXB with a BH primary \citep[e.g.,][]{bhevidence}, that has a mass of $2.3 M_{\odot} < M_{\rm BH} < 9.5 M_{\odot}$ \citep{infogx3394}. This system exhibits a 1.76\,day orbital period and has a K-type companion star. The distance to the source is still a matter of debate, with estimates ranging from $\sim$5--12\,kpc \citep[e.g.,][]{2004MNRAS.351..791Z,2019MNRAS.488.1026Z,infogx3394}. \source\ was first discovered by the Massachusetts Institute of Technology (MIT) X-ray detector on the \textit{Orbiting Solar Observatory 7} satellite in 1972, and was first detected in the radio band in 1994 by the Molonglo Observatory Synthesis Telescope at 843 MHz \citep{1994IAUC.6006....1S}. Undergoing numerous successful and hard-only outbursts over the past few decades \citep[so far 41\% from a total of 22 detected outbursts have been identified as hard-only;][]{watchdog}, \source\ is one of the best studied BH LMXBs \citep[e.g.,][]{2004MNRAS.351..791Z,2012AJ....143..130B}, particularly at both radio and X-ray wavelengths \citep[e.g.,][]{corbel}. Therefore, it is an ideal candidate to probe the disk--jet connection and identify accretion or jet signatures that may indicate whether an outburst will complete a full, successful outburst or not.

\begin{table*}
  \begin{center}
    \caption{X-ray evolution of \source\ during its 2017--2018 hard-only outburst. These parameters were determined from \Swift-XRT monitoring, where $N_{\rm H} = (0.56 \pm 0.02) \times 10^{22}$ cm$^{-2}$, which was tied and jointly fit across all epochs of this outburst. The total $\chi^2$ for the joint fit is 2129.36 with 2139 degrees of freedom. Errors are one sigma.}
    \label{xraydata}
    \begin{tabular}{cccccccc}
    \hline
          Date & MJD & ObsID & Count rate & $\Gamma$ & Normalisation & Unabsorbed flux & Hardness \\
          & & &  & & & (0.5 -- 10 keV) & (1.5 -- 10 keV / \\
          & & & & & & $\times 10^{-10}$ erg s$^{-1}$ cm$^{-2}$ & 0.5 -- 1.5 keV) \\
      \hline
      2017--09--29 & 58025 & 00032898146 & 1.20 $\pm$ 0.09 & 1.73 $\pm$ 0.15 & 0.012 $\pm$ 0.002 & 0.73 $\pm$ 0.06 & 2.61$^{+0.75}_{-0.70}$ \\
      2017--10--01 & 58027 & 00032898148 & 1.45 $\pm$ 0.09 & 1.20 $\pm$ 0.09 & 0.010 $\pm$ 0.001 & 1.1 $\pm$ 0.1 & 6.12$^{+1.15}_{-1.11}$  \\
      2017--10--03 & 58029 & 00032898149 & 1.15 $\pm$ 0.11 & 1.52 $\pm$ 0.14 & 0.018 $\pm$ 0.002 & 1.4 $\pm$ 0.1 & 3.64$^{+0.94}_{-0.88}$ \\
      2017--10--05 & 58031 & 00032898150 & 2.53 $\pm$ 0.10 & 1.18 $\pm$ 0.08 & 0.014 $\pm$ 0.001 & 1.64 $\pm$ 0.08 & 6.30$^{+0.96}_{-0.93}$ \\
      2017--10--07 & 58033 & 00032898151 & 4.24 $\pm$ 0.22 & 1.24 $\pm$ 0.09 & 0.026 $\pm$ 0.003 & 2.85 $\pm$ 0.19 & 5.74$^{+1.10}_{-1.06}$ \\
      2017--10--09 & 58035 & 00032898152 & 5.97 $\pm$ 0.25 & 1.23 $\pm$ 0.06 & 0.039 $\pm$ 0.003 & 4.3 $\pm$ 0.2 & 5.82$^{+0.73}_{-0.71}$  \\
      2017--10--17 & 58043 & 00032898153 & 10.71 $\pm$ 0.54 & 1.26 $\pm$ 0.08 & 0.062 $\pm$ 0.005 & 6.6 $\pm$ 0.4 & 5.54$^{+0.90}_{-0.86}$	 \\     
      2017--10--20 & 58046 & 00032898154 & 11.28 $\pm$ 0.48 & 1.30 $\pm$ 0.06 & 0.084 $\pm$ 0.006 & 8.45 $\pm$ 0.35 & 5.18$^{+0.65}_{-0.63}$	\\      
      2017--10--23 & 58049 & 00032898155 & 14.88 $\pm$ 0.15 & 1.58 $\pm$ 0.02 & 0.130 $\pm$ 0.003 & 9.4 $\pm$ 0.1 & 3.31$^{+0.15}_{-0.14}$ \\
      2017--10--25 & 58051 & 00032898158 & 15.76 $\pm$ 0.17 & 1.54 $\pm$ 0.03 & 0.097 $\pm$ 0.003 & 7.3 $\pm$ 0.1 & 3.50$^{+0.19}_{-0.18}$	 \\      
      2017--11--01 & 58058 & 00032898160 & 12.00 $\pm$ 0.14 & 1.45 $\pm$ 0.03 & 0.094 $\pm$ 0.003 & 7.8 $\pm$ 0.1 & 4.05$^{+0.20}_{-0.20}$ \\

      2018--01--20 & 58138 & 00032898161 & 5.93 $\pm$ 0.21 & 1.38 $\pm$ 0.06 & 0.041 $\pm$ 0.003 & 3.8 $\pm$ 0.2 & 4.53$^{+0.60}_{-0.58}$ \\      
      2018--01--25 & 58143 & 00032898162 & 3.84 $\pm$ 0.14 & 1.33 $\pm$ 0.09 & 0.029 $\pm$ 0.003 & 1.33 $\pm$ 0.09 & 4.94$^{+0.86}_{-0.82}$ \\     
      2018--01--30 & 58148 & 00032898163 & 2.23 $\pm$ 0.10 & 1.23 $\pm$ 0.07 & 0.015 $\pm$ 0.001 & 1.60 $\pm$ 0.08 & 5.75$^{+0.88}_{-0.85}$ \\      
      2018--02--04 & 58153 & 00032898164 & 1.32 $\pm$ 0.10 & 1.14 $\pm$ 0.12 & 0.007 $\pm$ 0.001 & 0.90 $\pm$ 0.08 & 6.73$^{+1.78}_{-1.69}$ \\     
      2018--02--07 & 58156 & 00032898165 & 1.47 $\pm$ 0.08 & 1.27 $\pm$ 0.09 & 0.007 $\pm$ 0.001 & 0.69 $\pm$ 0.05 & 5.40$^{+1.03}_{-0.99}$ \\      
      2018--02--09 & 58158 & 00032898166 & 1.11 $\pm$ 0.08 & 1.34 $\pm$ 0.1 & 0.008 $\pm$ 0.001 & 0.72 $\pm$ 0.05 & 4.87$^{+1.09}_{-1.04}$ \\
      2018--02--14 & 58163 & 00032898167 & 0.61 $\pm$ 0.04 & 1.52 $\pm$ 0.1 & 0.0047 $\pm$ 0.0004 & 0.36 $\pm$ 0.02 & 3.62$^{+0.65}_{-0.63}$ \\
      2018--02--27 & 58176 & 00032898170 & 0.28 $\pm$ 0.02 & 1.60 $\pm$ 0.15 & 0.0024 $\pm$ 0.0003 & 0.17 $\pm$ 0.01 & 3.21$^{+0.87}_{-0.82}$ \\

      \hline
    \end{tabular}
  \end{center}

\end{table*}

\section{Observations and analysis methods}

\label{sec:obs}

\subsection{X-ray analysis}

\subsubsection{\Swift-XRT}
\label{sec:swift}
X-ray observations of \source\ were taken by the Neil Gehrels \textit{Swift} Observatory X-ray telescope (XRT) throughout its 2017--2018 hard-only outburst. We used the \Swift-XRT online pipeline\footnote{\url{https://www.swift.ac.uk/user_objects/}} \citep{swiftXRT} to retrieve pile-up corrected spectral data. To compare behaviours between hard-only events of \source, we also downloaded observations taken during its 2013 hard-only outburst and 2018--2019 hard-only flare. We note that following this flaring event, while the source faded to almost quiescence, it did not completely return to quiescence before re-brightening in late-2019, going into a full outburst in 2019--2020 \citep{2019ATel13113....1R,2020ATel13447....1P}. Hence, we refer to the 2018--2019 hard-only event as a flare as opposed to a full outburst. For the 2017--2018 outburst, all \Swift-XRT data were obtained in photon counting (PC) mode. While data from the 2013 hard-only outburst and 2018--2019 hard-only flare were mostly obtained in the PC mode, some observations were taken in \Swift-XRT's windowed timing (WT) mode.

Since the focus of this work is the 2017--2018 outburst, only the epochs of this outburst will be discussed in detail, but we show all X-ray data and fitting parameters in the Appendix. The 2017--2018 \Swift-XRT data of this outburst consisted of 19 observations taken between 2017--09--29 (MJD 58025) and 2018--02--27 (MJD 58176), see Table~\ref{xraydata}. However, from the beginning of November 2017 until mid-January 2018, constraints due to the proximity of the Sun prevented \Swift-XRT from obtaining data of \source.

The data were analysed using \textsc{xspec} (version 12.8; \citealt{1996ASPC..101...17A}) from {\textsc{heasoft}} software package (version 6.21). The 0.5--10\,keV X-ray spectra were modelled with an absorbed power-law (\textsc{tbabs$\times$powerlaw}) within \textsc{xspec}. The {\tt tbabs} model component, making use of abundances from \citealt{2000ApJ...542..914W}, and photoionization cross-sections from \citealt{1996ApJ...465..487V}, was used to account for interstellar absorption. Tying the line of sight equivalent hydrogen absorption, \nh, across all epochs provided \nh\ = $(0.56 \pm 0.02) \times 10^{22}$ cm$^{-2}$. We also compared all results with \nh\ either fixed to literature values (\nh$_{\rm ,fixed} = 0.851 \times 10^{22}$ cm$^{-2}$; \citealt{2015ApJ...808..122F}), or left free (untied) for all epochs, finding our conclusions were not altered by the method that was used.

X-ray fluxes were determined using the convolution model \texttt{cflux} within \textsc{xspec}. The X-ray fluxes were measured for four different energy bands, namely 1--10 keV, 0.5--10 keV, 0.5--1.5 keV and 1.5--10 keV. 

For our X-ray analysis, the X-ray luminosity, $L_{\rm X}$, was calculated as $L_{\rm X} = 4 \pi F_{\rm X} D^2$, where $F_{\rm X}$ is the X-ray flux and $D$ is the source distance. In this work, we adopt a source distance of 8\,kpc \citep{2004MNRAS.351..791Z,2019MNRAS.488.1026Z,infogx3394}.

\subsubsection{\textit{MAXI}}

\begin{figure}
\centering
\includegraphics[width=1.03\columnwidth]{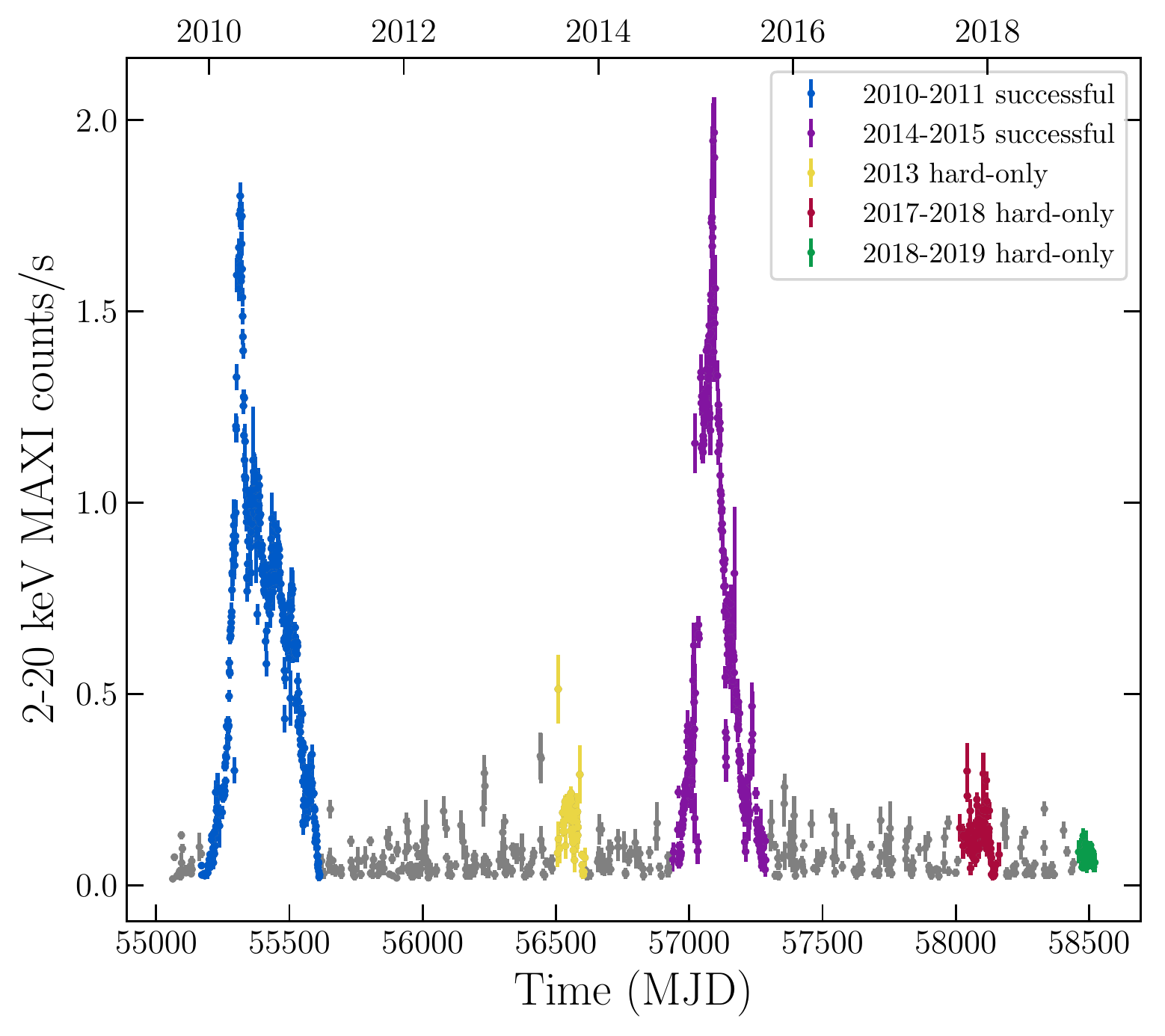}
\caption{2--20 keV \textit{MAXI} lightcurve of \source\ from August 2009 to February 2019, where five outbursts of this source have been highlighted in different colors. During this period, the three hard-only events were significantly less luminous than both successful outbursts and the 2018--2019 hard-only flare was especially faint in this X-ray band.}
\label{fig:MAXI_flux}
\end{figure}

\begin{figure*}
\centering
\includegraphics[width=1.0\textwidth]{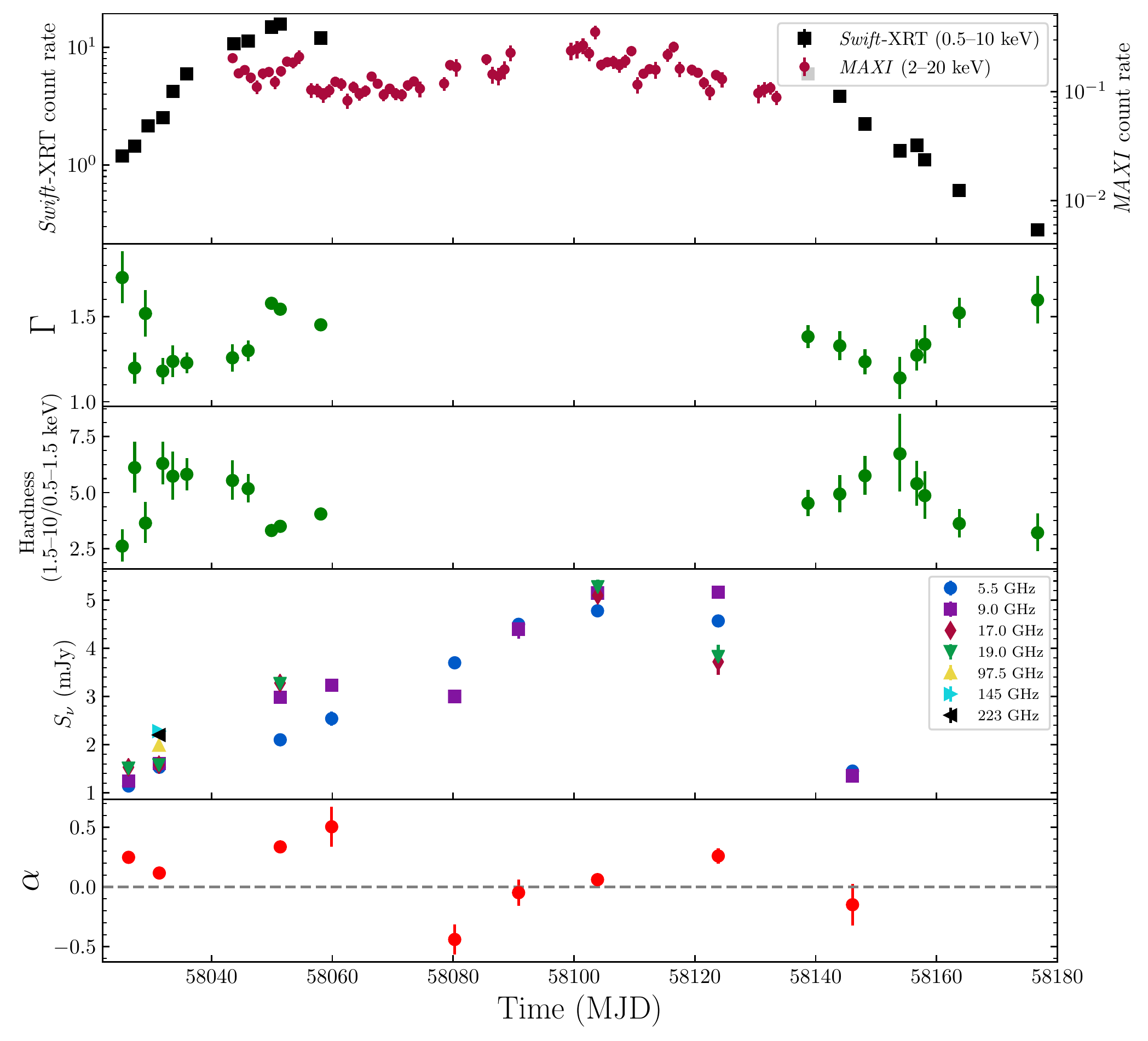}
\caption{X-ray and radio lightcurves of \source\ during its 2017--2018 hard-only outburst. (\textit{Top panel}): The 0.5--10 keV and 2--20 keV X-ray count rate evolution during the outburst, where the black squares are determined from \Swift-XRT and the red circles are from \textit{MAXI} when the source was Sun-constrained to \Swift-XRT (for clarity, only 5-$\sigma$ \textit{MAXI} detections are shown). (\textit{Second panel}): X-ray photon index, $\Gamma$, from the \Swift-XRT monitoring. (\textit{Third panel}): X-ray hardness, defined as 1.5--10 keV flux/0.5--1.5 keV flux. (\textit{Fourth panel}): Radio and (sub-)mm flux densities, $S_{\nu}$. (\textit{Fifth panel}): Radio spectral index, $\alpha$.  \source\ was sun constrained to \swift-XRT during the middle of this hard-only outburst.}
\label{combinedplot}
\end{figure*}

The Monitor of All-sky X-ray Image (\textit{MAXI}) X-ray telescope \citep{MAXIref} count rates (Figure~\ref{fig:MAXI_flux}) were obtained from the \textit{MAXI} website\footnote{\url{http://maxi.riken.jp/top/index.html}}. During times when \swift-XRT was sun constrained but radio observations were taken (MJDs~58080, 58090, 58103 and 58123), we used the \textit{MAXI} count rates to estimate the X-ray flux and luminosity (using WebPIMMS\footnote{\url{https://heasarc.gsfc.nasa.gov/cgi-bin/Tools/w3pimms/w3pimms.pl}}). To do this, we used the \nh\ found from our \swift-XRT analysis ($N_{\rm H} = (0.56 \pm 0.02) \times 10^{22}$ cm$^{-2}$), and assumed $\Gamma$ to be 1.5 -- 2.5, conservatively estimating the 1--10\,keV X-ray flux from the 2 -- 20\,keV {\it MAXI} count rate (Table~\ref{tab:MAXI_fluxes}). We do note that the unknown values of $\Gamma$ may make these fluxes unreliable, however, the range taken for $\Gamma$ provides a conservative estimate when taking into account the lack of dramatic changes to the X-ray count rate over this time (Figure~\ref{combinedplot}) and the minimal variation of $\Gamma$ during other hard-only outbursts from this source (during the brighter phase of the outburst, see Section~\ref{outburstcomparison}).

\begin{table}
  \begin{center}
    \caption{1--10\,keV X-ray fluxes and luminosities determined from the \textit{MAXI} count rates on dates where there was an ATCA radio observation and \source\ was sun constrained to \swift-XRT. The 1--10\,keV X-ray fluxes and luminosities were estimated from the 2--20\,keV \textit{MAXI} count rate by assuming an absorbed powerlaw with $N_{\rm H} = (0.56 \pm 0.02) \times 10^{22}$ cm$^{-2}$ and X-ray photon index of 1.5--2.5. To calculate the luminosity, we assumed a distance of 8\,kpc \citep{2019MNRAS.488.1026Z}.}
    \label{tab:MAXI_fluxes}
    \begin{tabular}{ccccc}
    \hline
          Date & MJD & Count rate & Flux & Luminosity \\
          & & (2--20\,keV) & (1--10\,keV) & (1--10\,keV)\\
          & & counts\,s$^{-1}$ & $\times 10^{-10}$ & $\times 10^{37}$\\
          & & & \,erg\,s$^{-1}$\,cm$^{-2}$ & erg\,s$^{-1}$\\

      \hline
      2017--11--23 & 58080 & 0.169 $\pm$ 0.031 & 24.8 $\pm$ 5.5 & 1.90 $\pm$ 0.42 \\
      2017--12--03 & 58090 & 0.139 $\pm$ 0.048 & 20.4 $\pm$ 4.5 & 1.56 $\pm$ 0.35 \\
      2017--12--16 & 58103 & 0.353 $\pm$ 0.052 & 52.0 $\pm$ 11.5 & 3.98 $\pm$ 0.88 \\
      2019--01--05 & 58123 & 0.142 $\pm$ 0.016 & 21.0 $\pm$ 4.6 & 1.61 $\pm$ 0.35 \\

       \hline
    \end{tabular}
  \end{center}
\end{table}

\subsection{ATCA radio data}
\label{sec:radio}
During the 2017--2018 outburst, GX~339$-$4 was observed 9 times with the Australia Telescope Compact Array (ATCA; PI: Russell, project code: C3057). Observations were carried out at central frequencies of 5.5 and 9 GHz on all dates and additionally at 17 and 19 GHz on 5 dates, see Table \ref{radiotable}. Each frequency pair (5.5 / 9 GHz or 17 / 19 GHz) was recorded simultaneously, with a bandwidth of 2\,GHz at each frequency band, which was comprised of 2048 channels of width 1\,MHz. Observations had 10 second integration times. PKS~1934--638 was used for primary flux and bandpass calibration, while the nearby source PKS~1646--50 was used for phase calibration. Data calibration was carried out following standard procedures from the Common Astronomy Software Application (\textsc{casa}, version 4.7.1; \citealt{casa}\footnote{\url{https://casaguides.nrao.edu/index.php?title=Main_Page}}). Each frequency band was imaged with natural weighting to maximize sensitivity. The signal to noise ratio was not high enough for reliable self-calibration. To determine the radio flux density, $S_{\nu}$, of the source for each epoch, we fit for a point source in the image plane, where we use a Gaussian with a full width half maximum (FWHM) equal to the synthesised beam of the observation. Errors on the absolute flux density scale include systematic uncertainties of 2\% for the 5.5/9\,GHz data\footnote{\url{https://www.narrabri.atnf.csiro.au/observing/users_guide/html_old_20090512/Flux_Scale_AT_Compact_Array.html}} 
(e.g.,
\newpage
\citealt{2011MNRAS.415.1597M};
\citealt{2012MNRAS.422.1527M})
and 4\% for the 17/19\,GHz data
\citep[see, e.g.,][]{2010MNRAS.402.2403M,2016ApJ...821...61P}. All radio flux densities are reported in Table \ref{radiotable}.

The radio luminosity ($L_{\rm r}$) was calculated using $L_{\rm r}$ = $4\pi \nu S_{\rm \nu} D^2$. Radio spectral indices ($\alpha$, where $S_{\nu} \propto \nu^{\alpha}$; \citealt{2001MNRAS.322...31F}) have been determined by Monte Carlo fitting using all the radio bands that were available for each date.

We also used \textsc{uvmultifit} \citep{2014A&A...563A.136M} to search for source intra-observational variability in the 5.5 and 9\,GHz observations. Unfortunately, the setup of the radio observations - where the 10 -- 15\,min scans of \source\ were taken sparsely during the full radio observation\footnote{Which was generally focused on a different source of interest.} - meant that the source was not densely sampled during the observation. Additionally, ATCA is a 6-element linear array meaning that the limited instantaneous uv-coverage did not allow variability shorter than 5-min intervals to be tested. For each time interval, we used \textsc{uvmultifit} to fit for a point source (which is a delta function in the uv-plane) at the target position (as well as all other sources in the field when detectable). Results for each variable epoch are provided in the Appendix, in Table~\ref{tab:radio_intra_obs_data} and Figure~\ref{fig:radiolc_var}.

\begin{table}
  \begin{center}
    \caption{ATCA radio and ALMA sub-mm flux densities and spectral indices ($\alpha$) in different GHz bands for the 2017--2018 outburst of \source. For the ATCA data, each band (5.5, 9.0, 17.0 and 19.0 GHz) has a $\pm 1$\,GHz range. Reported ALMA data have a bandwidth of 8\,GHz. Errors include systematic uncertainties.}
    \label{radiotable}
    \begin{tabular}{c c c c c}
    \hline
          Date & MJD & Frequency & Flux density & $\alpha$ \\
          & & (GHz) & (mJy) & \\
      \hline
      2017--09--30 & 58026 & 5.5 & 1.14 $\pm$ 0.06 & 0.25 $\pm$ 0.05 \\
       & & 9.0 & 1.24 $\pm$ 0.04 &  \\
       & & 17.0 & 1.53 $\pm$ 0.06 &  \\
       & & 19.0 & 1.51 $\pm$ 0.06 & \\
       2017--10--05 & 58031 & 5.5 & 1.53 $\pm$ 0.05 & 0.11 $\pm$ 0.03 \\
       & & 9.0 & 1.61 $\pm$ 0.05 &  \\
       & & 17.0 & 1.58 $\pm$ 0.06 &  \\
       & & 19.0 & 1.58 $\pm$ 0.06 &  \\
       & & 97.5 & 2.0 $\pm$ 0.1 & \\
       & & 145.0 & 2.28 $\pm$ 0.11 & \\
       & & 233.0 & 2.20 $\pm$ 0.11 & \\
       2017--10--25 & 58051 & 5.5 & 2.1 $\pm$ 0.1 & 0.34 $\pm$ 0.04 \\
       & & 9.0 & 2.98 $\pm$ 0.04 &  \\
       & & 17.0 & 3.28 $\pm$ 0.13 &  \\
       & & 19.0 & 3.27 $\pm$ 0.06 &  \\
       2017--11--02 & 58059 & 5.5 & 2.54 $\pm$ 0.15 & 0.50 $\pm$ 0.17 \\
       & & 9.0 & 3.23 $\pm$ 0.08 &  \\
       2017--11--23 & 58080 & 5.5 & 3.7 $\pm$ 0.1 & -0.44 $\pm$ 0.13 \\
       & & 9.0 & 3.00 $\pm$ 0.1 &  \\
       2017--12--03 & 58090 & 5.5 & 4.5 $\pm$ 0.1 & -0.05 $\pm$ 0.11 \\
       & & 9.0 & 4.4 $\pm$ 0.2 &  \\
       2017--12--16 & 58103 & 5.5 & 4.78 $\pm$ 0.06 & 0.06 $\pm$ 0.02 \\
       & & 9.0 & 5.15 $\pm$ 0.07 &  \\
       & & 17.0 & 5.10 $\pm$ 0.15 &  \\
       & & 19.0 & 5.28 $\pm$ 0.15 &  \\
       2018--01--05 & 58123 & 5.5 & 4.57 $\pm$ 0.05 & 0.26 $\pm$ 0.06 \\
       & & 9.0 & 5.17 $\pm$ 0.07 &  \\
       2018--01--27 & 58146 & 5.5 & 1.45 $\pm$ 0.08 & -0.15 $\pm$ 0.18 \\
       & & 9.0 & 1.35 $\pm$ 0.08 &  \\
       \hline
    \end{tabular}
  \end{center}
\end{table}

\subsection{ALMA (sub)-millimetre data}
The Atacama Large Millimetre/Sub-Millimetre Array (ALMA) observed GX 339--4 (PI: Tetarenko, project code: 2017.1.00864.T) on 2017 Oct 05 (21:23:05.0 -- 23:04:36.3 UTC; MJD 58031.8910 -- 58031.9615). Data were taken sequentially in Bands 3, 4, and 6, at central frequencies of 97.5, 145, and 233\,GHz, respectively. The ALMA correlator was set up to yield $4\times2$ GHz wide base-bands at each frequency band. During our observations, the 12m array was in its Cycle 5 C41-9 configuration, with 41 antennas, spending $\sim$ 8.1/9.8/15.7 min total on the target source in Bands 3, 4, and 6, respectively. The median precipitable water vapour (PWV) during the observations was 0.78, 0.71, and 0.67 mm for the Bands 3, 4, and 6 observations, respectively. All of the data were reduced and imaged within \textsc{casa} (version 5.1.1; \citealt{casa}), using standard procedures outlined in the \textsc{casa}Guides for ALMA data reduction\footnote{\url{https://casaguides.nrao.edu/index.php/ALMAguides}}. We used J1617--5848/J1427--4206 as bandpass \& flux calibrators, J1650--5044 as a phase calibrator, and J1631--5256 as a check source, for all the observations. To image the continuum emission, we performed multi-frequency synthesis imaging on the data using the \texttt{tclean} task within \textsc{casa}, with natural weighting to maximize sensitivity. Flux densities of the source were then measured by fitting a point source in the image plane (using the \texttt{imfit} task). All ALMA sub-mm flux densities are recorded in Table~\ref{radiotable}. Systematic errors were applied to the absolute flux density calibration of the ALMA data, where an uncertainty of 5\% is expected for ALMA bands
$<$350\,GHz\footnote{\url{https://almascience.eso.org/documents-and-tools/latest/documents-and-tools/cycle8/alma-technical-handbook}}. We also explored the ALMA sub-mm observations for intra-observational variability. Results are discussed in Section~\ref{sec:intra_obs_variability}, with light curves and the data points provided in Appendix~\ref{sec:appendix_variability}.

\section{Results}
\label{sec:results}

\subsection{X-ray lightcurves and spectral evolution}

During the 2017--2018 outburst of GX 339$-$4, the X-ray lightcurve shows a roughly single rise and decay evolution with an exponential decay (Figure~\ref{combinedplot}, top panel). \swift-XRT monitoring showed a peak 0.5--10\,keV X-ray flux of $\approx 9.4 \times 10^{-10}$\,\ergscm\ at MJD 58049. During the Sun constraints to \swift, the source brightened, reaching a peak 2--20\,keV flux of $\approx 5.2 \times 10^{-9}$\,\ergscm\ at MJD~58103, which corresponds to a 0.5--10\,keV flux of $\approx 7 \times 10^{-9}$\,\ergscm\ (assuming $\Gamma = 2$). 

During the early stages of our X-ray monitoring, the X-ray photon index ($\Gamma$) was variable, where we initially detected a softer X-ray photon index ($\Gamma = 1.73 \pm 0.15$), which then hardened (to $\Gamma = 1.20 \pm 0.09$), before softening again (to $\Gamma = 1.52 \pm 0.14$) and then re-hardening (to $\Gamma = 1.18 \pm 0.08$) once again over the space of about a week (Figure \ref{combinedplot}, panel 2). After this variable behaviour, $\Gamma$ then followed a generally standard pattern of evolution for a hard-only outburst, where it only marginally steepened as the outburst brightened, reaching a $\Gamma$ of $\approx$ 1.58 at the peak X-ray flux observed by \swift-XRT. However, instead of brightening and softening further, the outburst began to fade and the X-ray spectrum hardened.

Once the X-ray flux began to decrease, $\Gamma$ also started decreasing in a reverse pattern to its behaviour during the rise phase of the outburst, such that $\Gamma$ hardened to $\approx 1.15$ (Figure~\ref{combinedplot}). Then, similar to the rise phase, at our lowest observed X-ray fluxes (below $9 \times 10^{-11}$\,\ergscm) the X-ray photon index then softened progressively as the source faded, evolving to $\Gamma \approx 1.6$ by the end of our monitoring. An analysis of the X-ray hardness, where we use the ratio between the 1.5--10 keV and 0.5--1.5 keV X-ray flux, showed a similar pattern of behaviour for the 2017--2018 outburst (Figure \ref{combinedplot}, panel 3).

\subsection{Radio/sub-mm lightcurves}

\begin{figure}
\centering
\includegraphics[width=1.0\columnwidth]{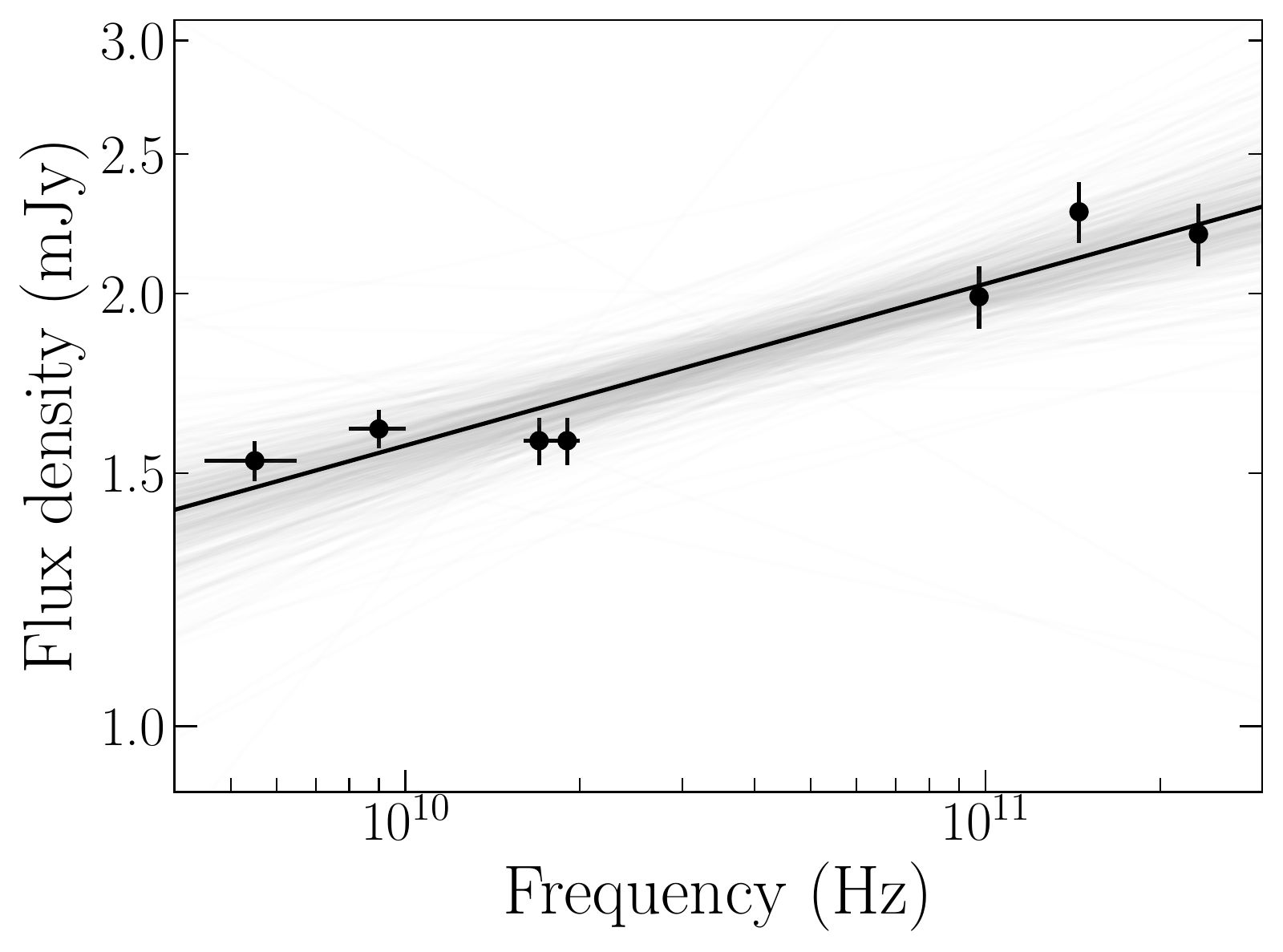}
\caption{The radio to sub-mm spectrum taken on 2017-10-05 (MJD~58031). Despite the $>$0.5\,day difference between the ATCA and ALMA observations, and the observed source variability, these broadband data are reasonably well represented by a single power law with a spectral index $\alpha$ of $0.11 \pm 0.03$ (black line, with errors shown by the fainter grey lines), indicatingg emission an optically-thick compact jet.}
\label{fig:oct5_spectrum}
\end{figure}

In our 5-month long radio/sub-mm monitoring of GX 339--4, we observed the integrated radio/sub-mm flux densities vary between $\sim1-5$ mJy. Long-term light curves revealed a single rise and decay phase (Figure \ref{combinedplot}, panel 4), where the spectral index ($\alpha$) remained flat to inverted (and well represented by a single power law; Figure~\ref{fig:oct5_spectrum}), except for our radio observation taken on MJD~58080 (Figure~\ref{combinedplot}, panel 5), which displayed a relatively steep radio spectral index ($\alpha = -0.44 \pm 0.13$). However, as shown below (Section~\ref{sec:intra_obs_variability} and Figure~\ref{fig:radiolc_var}), we see a declining trend in the flux density of the radio bands throughout this observation. This is suggestive that the steep spectral index from this epoch could be the result of compact jet variability (where we might expect a delay between variability features of up to tens of minutes between these radio bands, towards the lower frequency band; \citealt{2019MNRAS.484.2987T}). In addition, we do not expect transient ejecta (which tend to show a steep spectrum) to be launched at such a low X-ray luminosity and hard X-ray spectrum, where there does not appear to be any sudden X-ray changes. Therefore, the radio and sub-mm results are consistent with emission from a compact, partially self-absorbed synchrotron jet. This emission remained on for the entire outburst, with no indication of the jet emission being quenched in our observations.

\subsubsection{Intra-observational variability}
\label{sec:intra_obs_variability}

Compact jets have been found to be highly variable on short ($<1$ day) timescales (e.g., \citealt{2019MNRAS.484.2987T}). As such, we opted to search for intra-observational variability in our radio through sub-mm data of GX 339--4. Figure~\ref{fig:radiolc_var} displays intra-observation light curves from the ATCA radio data sampled on 5-min timescales, while Figure~\ref{fig:submmlc_var} displays intra-observation light curves from the ALMA sub-mm data sampled on 30-sec timescales.

To characterize the amplitude of any variability and compare between frequency bands, we use the fractional root mean square (RMS) statistic,
\begin{equation}
F_{\rm var}=\sqrt{\frac{S^2-\bar{\sigma}_{\rm err}^2}{\bar{x}^2}},
\end{equation}
where $\bar{x}$ represents the weighted mean of the flux measurements, the sample variance $S^2=\frac{1}{N-1}\sum_{i=1}^{N}(x_i-\bar{x})^2$, and the mean square measurement error $\bar{\sigma}_{\rm err}^2=\frac{1}{N}\sum_{i=1}^{N}\sigma_{\rm err,i}^2$ \citep{akr96,vau03,sad06}. We consider $F_{\rm var}< 15$\% as not significantly variable, $15\%<F_{\rm var}< 30$\% as mildly variable, and $F_{\rm var}> 30$\% as highly variable \citep[as in][]{2019MNRAS.482.2950T}.

In the ALMA sub-mm data, we detect clear short-timescale variability, in the form of structured, small-scale flaring episodes across all three bands sampled. 
For example, the largest flare detected at 230 GHz ($\sim$ 21:40 UT) is symmetric in shape, and rises to an amplitude of $\sim 5$ times the base flux density level of $\sim$1\,mJy, over a timescale of $\sim 2$ min (corresponding to a brightness temperature of $\sim 10^8$ K, consistent with other synchrotron events from LMXBs; \citealt{2015MNRAS.446.3687P}).
We measure fractional RMS values of $21.6\pm0.1$ \%, $35.9\pm0.2$ \%, and $35.4\pm0.1$ \% for the 97, 145, and 230\,GHz bands, respectively, indicating a mildly variable to highly variable source at these frequencies.
To ensure that the sub-mm variations we observe from \source\ are intrinsic to the source, and not due to atmospheric or instrumental effects, we also extracted intra-observation light curves for our check source. We find that the check source shows relatively constant flux densities throughout our observations in all the bands (with any variations present being $<10$\% of the average flux density), thus we are confident that the variations we observed from \source\ are an accurate representation of the rapidly changing intrinsic flux density of the source. 

In the ATCA radio data, we also detected short-timescale variability. Due to the low-instantaneous uv-coverage of the ATCA interferometric array, we were only able to explore variability down to 5-minute timescales. Additionally, the observing setup meant that only a few 10-minute scans of \source\ were taken spread out within a much longer observation. Therefore, our observations do not, and are not as sensitive to, detecting structured flares similar to those we observed within the ALMA sub-mm data. However, we do observe increasing/decreasing trends both within (e.g.,\ 2017-10-05) and throughout some observations (e.g.,\ 2017-09-30, 2017-11-23). This behaviour suggests that small-scale flaring activity, similar to that seen in the sub-mm bands, may have also been occurring in the radio bands, but we were not able to sample the source on short enough time-scales to resolve the flares. We measure fractional RMS values ranging from $1.9-15.8$ \% at 5\,GHz and $2.4-17.0$ \% at 9\,GHz. While this may indicate a mildly variable source at times, the source was too faint for self-calibration and, therefore, the results may suffer from phase decorrelation or gain drifts. For example, when comparing this variability to a check source present in the ATCA field of view\footnote{Note that we are only able to detect the check source for the three earliest ATCA observations due to the array configuration changing from a very compact to much more extended configuration on 2017-11-01.}, we find the check source can at times be just as variable as GX 339--4. This, plus the low level of variability, suggests that the radio variations we observe from GX 339--4 may not be completely intrinsic to the source, with possible contributions from atmospheric or instrumental effects.

Comparing the variability properties between the radio and sub-mm bands, we see a pattern of higher variability amplitudes at higher electromagnetic frequencies. This trend is consistent with what we might expect from compact jet emission, as the (sub-)mm emission originates from a region close to the base of the jet (with a smaller cross-section), while the radio emission originates from a region further out in the jet flow (with a large cross-section). Imaging the ALMA data on finer frequency scales (in 2\,GHz sub-bands for all bands) shows that the in-band ALMA spectral indices were flat/inverted throughout the observation (Table~\ref{tab:ALMA_inband}). While these results were time-averaged, they strongly support the suggestion that the observed flux density variations arose from a variable compact jet.

\begin{table}
  \begin{center}
    \caption{ALMA flux densities from each 2\,GHz sub-band. The flat-to-inverted (sub-)mm spectrum indicates optically-thick emission from a variable compact jet.}
    \label{tab:ALMA_inband}
    \begin{tabular}{cc}
    \hline
          Frequency & Flux Density \\
          ($\pm$2) GHz & mJy \\

      \hline
      90.5 & 2.0$\pm$0.1 \\
      92.5 & 1.9$\pm$0.1 \\
      102.5 & 2.1$\pm$0.1 \\
      104.5 & 2.1$\pm$0.1 \\   

      138 & 2.1$\pm$0.1 \\
      140 & 2.3$\pm$0.1 \\
      150 & 2.5$\pm$0.1 \\
      152 & 2.5$\pm$0.1 \\   

      224 & 2.2$\pm$0.1 \\
      226 & 2.2$\pm$0.1 \\
      240 & 2.3$\pm$0.1 \\
      242 & 2.3$\pm$0.1 \\

       \hline
    \end{tabular}
  \end{center}
\end{table}

\section{Discussion}
\label{sec:discussion}

\subsection{Source brightness}

The 2017--2018 hard-only outburst of \source\ was well monitored in the radio and X-ray bands at low X-ray luminosities during the rise phase due to the early discovery of the outburst from regular optical monitoring \citep{2017ATel10797....1R}\footnote{See \citealt{2019AN....340..278R} for full details on the X-ray Binary New Early Warning System (XB-NEWS).}. As such, our X-ray and radio monitoring began at an X-ray luminosity of $\approx$ 5.6 $\times$ 10$^{35}$\,erg\,s$^{-1}$, corresponding to an Eddington luminosity, $L_{\rm Edd}$, of $\sim$ 5.8 $\times$ 10$^{-4}$ (assuming a distance of 8\,kpc and $M_{BH} = 7.8\,M_{\odot}$, which is the peak of the best-fit mass distribution of the BH LMXB population; \citealt{2010ApJ...725.1918O}, where $L_{\rm Edd} = 1.26$ $\times$ $10^{38}$ $(M_{BH} / M_{\odot})$\,erg\,s$^{-1}$), and our first radio detection occurred at $L_{\rm r} \approx 4.2 \times 10^{29}$\,erg\,s$^{-1}$. At the X-ray peak of the outburst, \source\ reached an X-ray luminosity of $\sim$ 4 $\times$ 10$^{37}$ erg s$^{-1}$, corresponding to $L_{\rm Edd} \sim 0.04$. See also \citet{2020ApJ...899...44W} for a discussion on the X-ray luminosity determined with \textit{NICER} (while our results generally agree, their results are taken from observations averaged over multiple days, making a direct comparison difficult). In the radio band, we measured a maximum luminosity of $\sim$ 2 $\times$ 10$^{30}$\,erg\,s$^{-1}$.  

As expected, at its X-ray and radio peak, this outburst was significantly fainter than luminosities reached from successful outbursts of \source\ \citep{corbel}, as well as those from other systems, where successful outburst luminosities usually exceed $10^{38}$\,erg\,s$^{-1}$ \citep{watchdog}. In comparison to other hard-only outbursts from \source, the 2017--2018 outburst reached similar X-ray luminosities to other hard-only outbursts, within a factor of a few. The observed \source\ luminosities were also in broad agreement with typical peak luminosities from the full sample of hard-only outbursts from other systems ($L_{\rm X} \sim 10^{35}$--$10^{37}$\,erg\,s$^{-1}$; \citealt{watchdog}), albeit on the higher end of that range. Additionally, the 2017--2018 peak X-ray luminosity occurred at a similar luminosity to the expected hard state to HIMS transition luminosity for a large sample of BH LMXBs ($\geq 0.03 L_{\rm Edd}$; \citealt{2003A&A...409..697M,2010MNRAS.403...61D,2013ApJ...779...95K,2019MNRAS.485.2744V}), but below the typical transition luminosities observed from \source\ ($\sim 0.04 - 0.07 L_{\rm Edd}$; \citealt{2019MNRAS.485.2744V}).

\begin{figure}
\centering
\includegraphics[width=0.95\columnwidth]{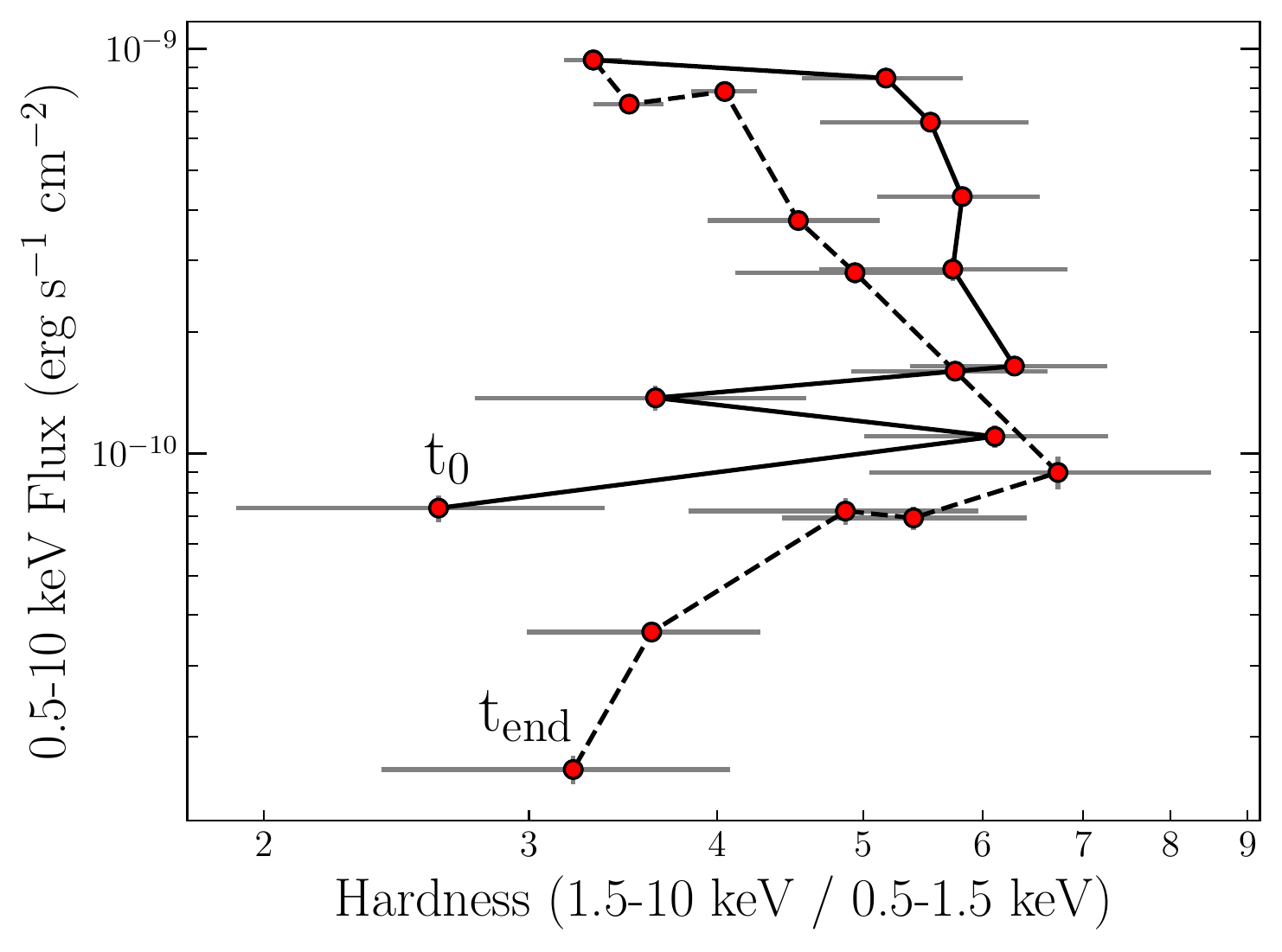}
\caption{Hardness-intensity diagram of the 2017--2018 outburst, calculated using \Swift-XRT data. Here, the hardness is defined as the 1.5-10 keV flux over the 0.5-1.5 keV flux. The first \swift-XRT observation is marked with $t_{0}$ and the last observation with $t_{\rm end}$. The solid black line represents the rise phase of the outburst, while the dashed black line characterises the decay. The outburst showed a softer X-ray spectrum at lower X-ray luminosities. Additionally, the source did not evolve to the soft X-ray spectral state in the outburst, but did soften marginally close to the outburst peak.}
\label{HIDgx3394}
\end{figure}

\subsection{X-ray spectral evolution}
\label{phoindexsection}

During the early stages of our monitoring, at the lowest observed (1--10\,keV) X-ray luminosities of the rise phase ($\lesssim1.3 \times$ 10$^{-3}$ \lEdd), \source\ exhibited a relatively soft X-ray spectrum. As the source brightened, the X-ray spectrum hardened (Figure~\ref{combinedplot}). This evolution was demonstrated by the variable changes in the source hardness (Figure~\ref{HIDgx3394}) and the X-ray photon index, where $\Gamma$ evolved back and forth between $\sim 1.7$ and $\sim 1.2$ over a few days; see Table \ref{xraydata}. Following this erratic evolution, at X-ray luminosities of $\gtrsim 1.3 \times 10^{-3}$ \lEdd, $\Gamma$ increased progressively to $\approx 1.6$ at the peak flux of the outburst. As the outburst faded, $\Gamma$ hardened. Similar to the early stages of the outburst, at the lowest observed X-ray luminosities at the end of the outburst ($\lesssim$ 7 $\times$ 10$^{-4}$ \lEdd), the X-ray spectrum again softened, although this decay phase evolution was much more gradual than during the rise phase. While the gradual softening and hardening of the X-ray spectrum during the bright phase of an outburst is standard, spectral hardening at low X-ray luminosities in the rise phase is generally not observed, although, as we discuss below, this is likely due to a lack of monitoring at such low X-ray luminosities early in the outburst.

In their quiescent state, BH LMXBs show softer X-ray spectra, such that $\Gamma \approx 2$ \citep[e.g.,][]{plotkin2013,2014MNRAS.441.3656R}. \source\ was not in quiescence during our monitoring (our observations begin at an X-ray luminosity of $\sim 10^{-3}\,L_{\rm Edd}$, well above the expected quiescence level of $\sim 10^{-5}\,L_{\rm Edd}$ for \source\ \citep{plotkin2013}. It has been suggested that at low X-ray luminosities, the X-ray spectrum could be dominated by emission from the base of the jet \citep[e.g.,][]{2005ApJ...635.1203M,2009MNRAS.398.1638M,2019MNRAS.485.3696C,2019MNRAS.482.4798L}, where changes to the observed spectrum could arise from a change in the location of the synchrotron cooling break, $\nu_{\rm cool}$, which represents the frequency at which the radiation timescales of the synchrotron emission are shorter than the dynamical time scales of the emitting electrons\footnote{The position and evolution of the cooling break are very poorly understood; its location has been inferred in the X-ray band at $\sim 10^{-3}$\,\lEdd, in the UV-band at $\sim 10^{-5}$\,\lEdd\ when the source enters quiescence \citep{2012ApJ...753..177P,2013MNRAS.429..815R,2013MNRAS.434.2696S}, and possibly in the optical band at the highest X-ray luminosities (around the transition from the hard state to the soft state; \citealt{2014MNRAS.439.1390R}).}. While the shape of the X-ray spectrum alone may suggest that the low-luminosity X-ray emission could have been dominated by the un-cooled optically-thin emission from the jet\footnote{Although we would expect $\Gamma$ to be close to typical optically-thin synchrotron spectral indices (such that $\Gamma \sim 1.5 - 1.7$, where $\alpha_{\rm thin} = 1 - \Gamma$). Then, when $\nu_{\rm cool}$ would be below the X-ray band, the X-ray spectrum should appear even steeper (steepening by a half, such that $\Gamma \sim 2 - 2.2$), which we did not observe.}, using broadband \textit{NuSTAR} X-ray observations, \citet{2019ApJ...885...48G} detected strong signatures of X-ray reprocessing on 2017-10-02 (see also \citep{2020ApJ...899...44W}), which are not expected if the X-ray emission is synchrotron in origin \citep{2004ApJ...609..972M}. Instead, \citet{2019ApJ...885...48G} proposed that the reprocessing originated in an optically-thick medium, presumably the accretion disk. In such a case, changes to the optical depth of the disk would have produced observable changes in the X-ray spectrum (such that an increasing optical depth would enhance the high-energy `hard' X-ray photons) at luminosities similar to our observations \citep[e.g.,][]{2020ApJ...889L..18Y}. As such, rapid changes to the optical depth and geometry \citep{2020ApJ...899...44W} of the accretion disk could result in the erratic spectral evolution we observed during the low-luminosity stage of the outburst rise. In addition, a further argument against synchrotron cooled X-rays is that they are expected to yield a steeper correlation between the radio and X-ray luminosities \citep{2005ApJ...629..408Y}, which we did not observe, instead detecting a shallower correlation (see Section~\ref{sec:lrlx} for results and further discussion).

Alternatively, similar changes to the X-ray spectral shape could also be observed from an optically-thin flow in a RIAF \citep[e.g.,][]{1995ApJ...452..710N,1997ApJ...489..865E,2014ARA&A..52..529Y}. In such a case, if the X-ray emission arises via inverse Compton in the outer disk or synchrotron self-Compton in the inner flow, then both gradual and erratic changes in the X-ray spectrum can be produced simply by changes to the optical depth of the inverse Compton scatterings \citep[e.g.,][]{1997ApJ...489..865E}. As such, the X-ray observations imply that the X-ray spectral changes were a result of an evolving optical depth of the accretion flow, whether it is optically-thick or thin.

We do note that during the low-luminosity decay phase (from 2018-02-04 and on-wards) our monitoring does not allow us to identify the origin of the hard X-ray emission. \citet{2019ApJ...885...48G} find signatures of X-ray reprocessing on 2018-01-30, but there were no high spectral resolution X-ray observations after those times. Therefore, it might be possible that the hard X-rays were dominated by emission from the base of the jet at later times.

\begin{figure}
\centering
\includegraphics[width=1.0\columnwidth]{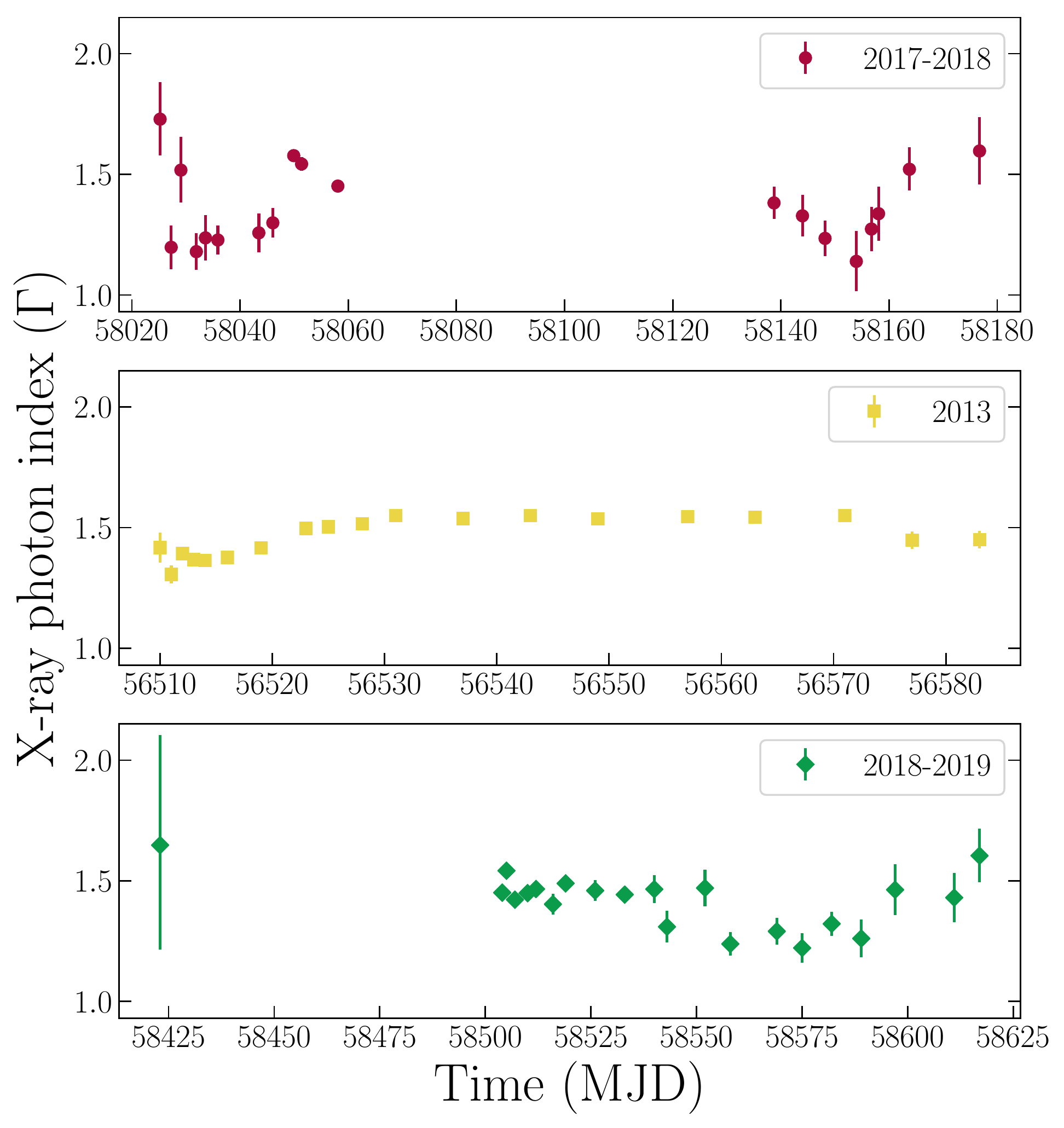}
\caption{The evolution of the X-ray photon index over time for three different hard-only GX~339$-$4 outbursts or flaring events. The top panel shows our 2017--2018 outburst. The second and third panels show the 2013 hard-only outburst and 2018--2019 hard-only flare. For all panels the range of photon indices are kept the same for clearer comparison. Note that the early rise part of the 2018--2019 flare was not observed by \Swift-XRT due to Sun constraints. The X-ray photon index evolution of the 2017--2018 outburst appears somewhat different to the other two failed outbursts shown here, where $\Gamma$ was initially steep, before flattening as the outburst progressed. Additionally, at the end of the outburst, $\Gamma$ increased.}
\label{phoindex_comparison}
\end{figure}

\begin{figure}
\centering
\includegraphics[width=1.0\columnwidth]{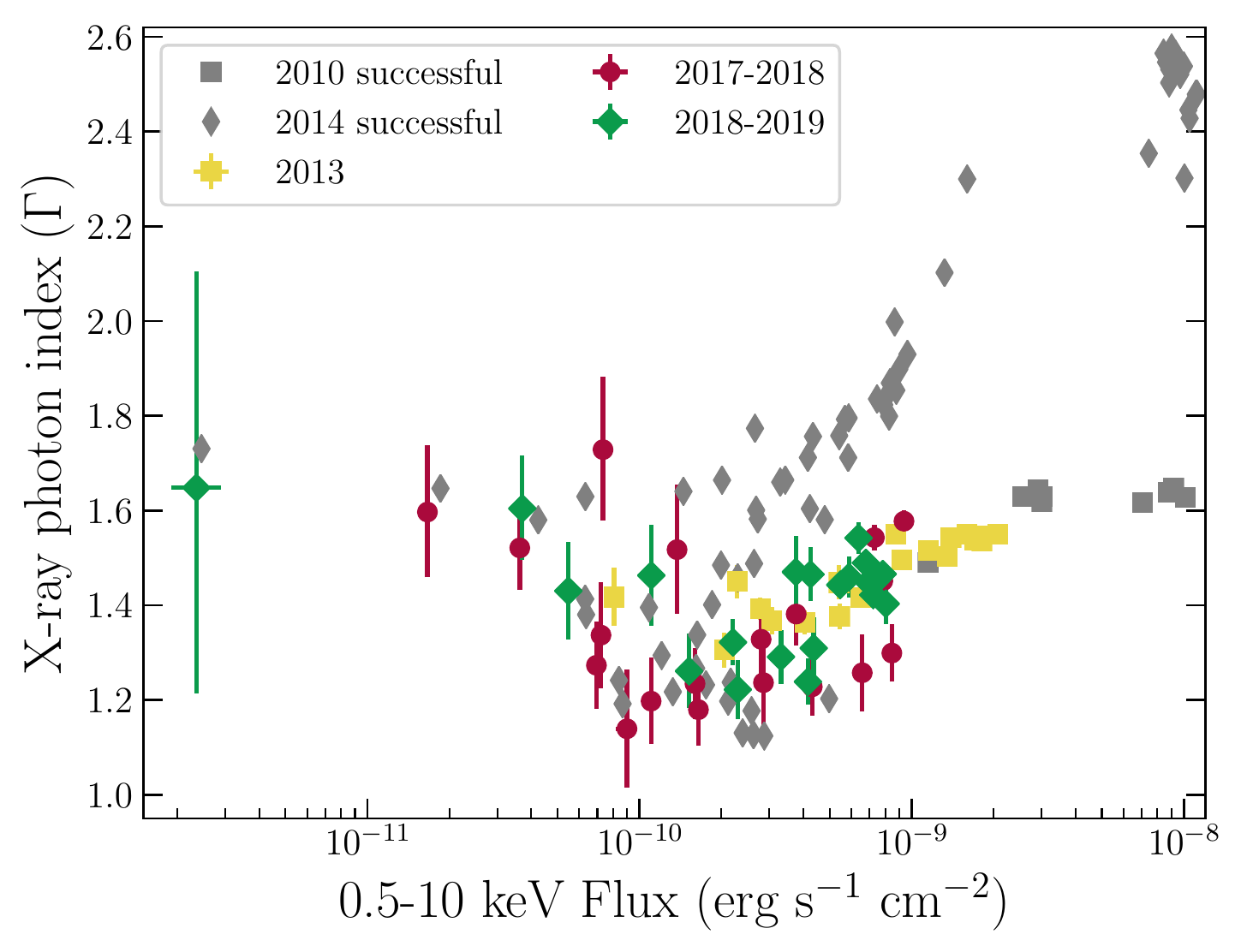}
\caption{The X-ray photon index vs flux for the 2013 (yellow) and 2017--2018 (red) hard-only outbursts, and the 2018--2019 hard-only flare (green) of \source, compared to \swift-XRT data from the 2010--2011 and 2014--2015 successful outbursts (grey). The evolution during the rise (lower path) and the decay (upper path) appear as two different tracks. This highlights the similarity (during the rise phase) and difference (during the decay phase) in evolution of the X-ray photon index during hard-only and successful outbursts.}
\label{phoindex_flux}
\end{figure}

\subsubsection{Comparison with other outbursts}
\label{outburstcomparison}

A simple comparison of the X-ray spectral evolution against the 2013 hard-only outburst and 2018-2019 hard-only flare from \source\ (Figure \ref{phoindex_comparison}) does not show a softer X-ray spectral evolution compared to the 2017--2018 outburst. However, the spectral softening and variability during the earliest stages of our monitoring do not appear to have been observed in these other hard-only outbursts, although, as shown below, this is due to missing data early on in these outbursts. 

To best compare the evolution of the X-ray photon index between outbursts (both hard-only and successful), we explored how $\Gamma$ evolved with X-ray flux for the 2013 and 2017--2018 hard-only outbursts and 2018--2019 hard-only flare, and a sample of two successful outbursts (Figure~\ref{phoindex_flux}). Such a comparison shows clear similarities despite the differences in cadence and luminosity of the observations. At lower X-ray luminosities as the flux increases, $\Gamma$ first decreases before increasing at higher X-ray fluxes, displaying a regular ``V''-shaped pattern. The only clear difference between the two outburst types is the hysteresis of $\Gamma$ with X-ray flux: hard-only outbursts appear to traverse the same path during both the rise and decay phases of the outburst, showing little to no hysteresis in $\Gamma$ (Figure~\ref{phoindex_flux}), while successful outbursts appear to follow two different tracks - a harder (lower) track during the rise, and a softer (higher) track during the decay, before re-joining as the source moves to X-ray fluxes $\lesssim 10^{-10}$\,erg\,s$^{-1}$\,cm$^{-2}$. The V-shaped pattern between the X-ray flux and $\Gamma$ (Figure~\ref{phoindex_flux}) at similar luminosities is commonly observed in BH LMXBs \citep[see, e.g.,][figure 1]{2008ApJ...682..212W}. The inflection point, where the source switches from X-ray spectral hardening to softening with increased X-ray luminosity, has been attributed to a transition from a RIAF (at bolometric luminosities of $<$ 0.01\,$L_{\rm Edd}$) to a standard accretion disk ($L_{\rm Bol} >$ 0.01\,$L_{\rm Edd}$; \citealt{2008ApJ...682..212W}). 

However, other possibilities for the inflection point have also been suggested, such as the point where emission becomes dominated by the jet, or where reprocessed photons begin to dominate in an out-flowing coronal model (e.g.,\ \citealt{2011MNRAS.417..280S}).

\begin{figure}
\centering
\includegraphics[width=1.0\columnwidth]{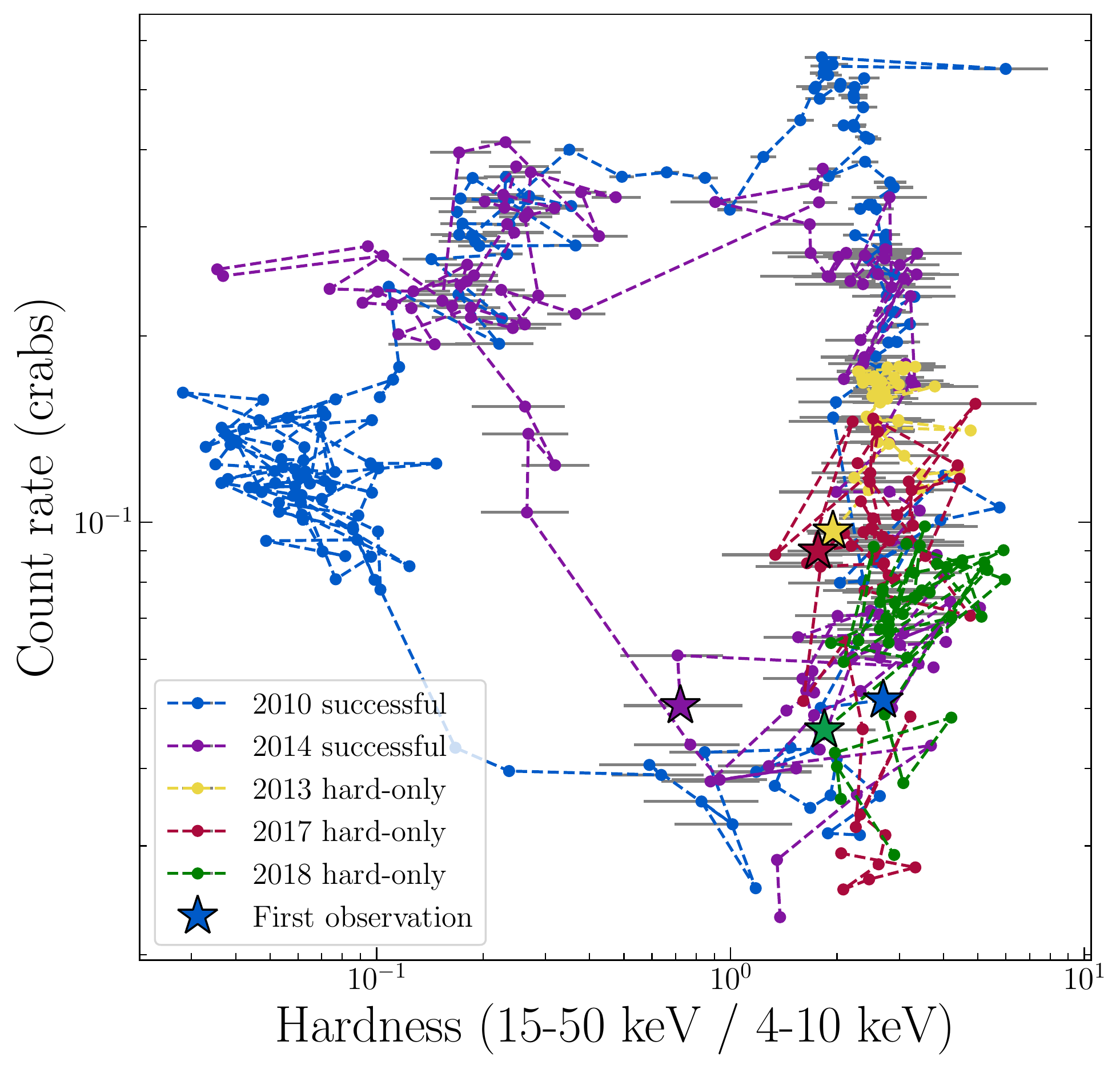}
\caption{Hardness-intensity diagram for three hard-only events and two successful outbursts from \source, using data from the \textit{MAXI} X-ray telescope and \Swift-BAT. Here, we define hardness as the 15-50 keV flux (from \Swift-BAT) over the 4--10 keV flux (from \textit{MAXI}). The larger stars correspond to the first observations of each outbursts. For clarity, we show error bars in grey. This figure indicates that there does not appear to be any significant differences in the \textit{MAXI} and \Swift-BAT data between hard-only and successful outbursts from \source\,during their rise phases.}
\label{HID_MAXI_all}
\end{figure}

To identify any deviation in source behaviour during the Sun constraint, we also compared the X-ray hardness evolution with X-ray intensity for both hard-only and successful outbursts of this source using \textit{MAXI} and \Swift-BAT observations (Figure~\ref{HID_MAXI_all}). Our results show that despite showing some minor hysteresis in the \swift-XRT data, the 2017--2018 outburst hardness evolution did not appear to significantly deviate from other outbursts of this source. This applies to the rise phase of successful outbursts and complete hard-only outbursts. 

Therefore, the 2017--2018 X-ray outburst of \source\ seems to follow a relatively typical evolution when compared to the X-ray behaviour of both hard-only and successful outbursts, with no preliminary indicators that may identify whether an outburst is successful or not. A similar result was also found during detailed near-infrared (NIR) and optical studies of outbursts from BH LMXBs \citep{2020A&A...638A.127K}, where no indicators in the NIR/optical colours or magnitude were identified that might allow for an identification of a hard-only or successful outburst.

\begin{figure}
\centering
\includegraphics[width=1.0\columnwidth]{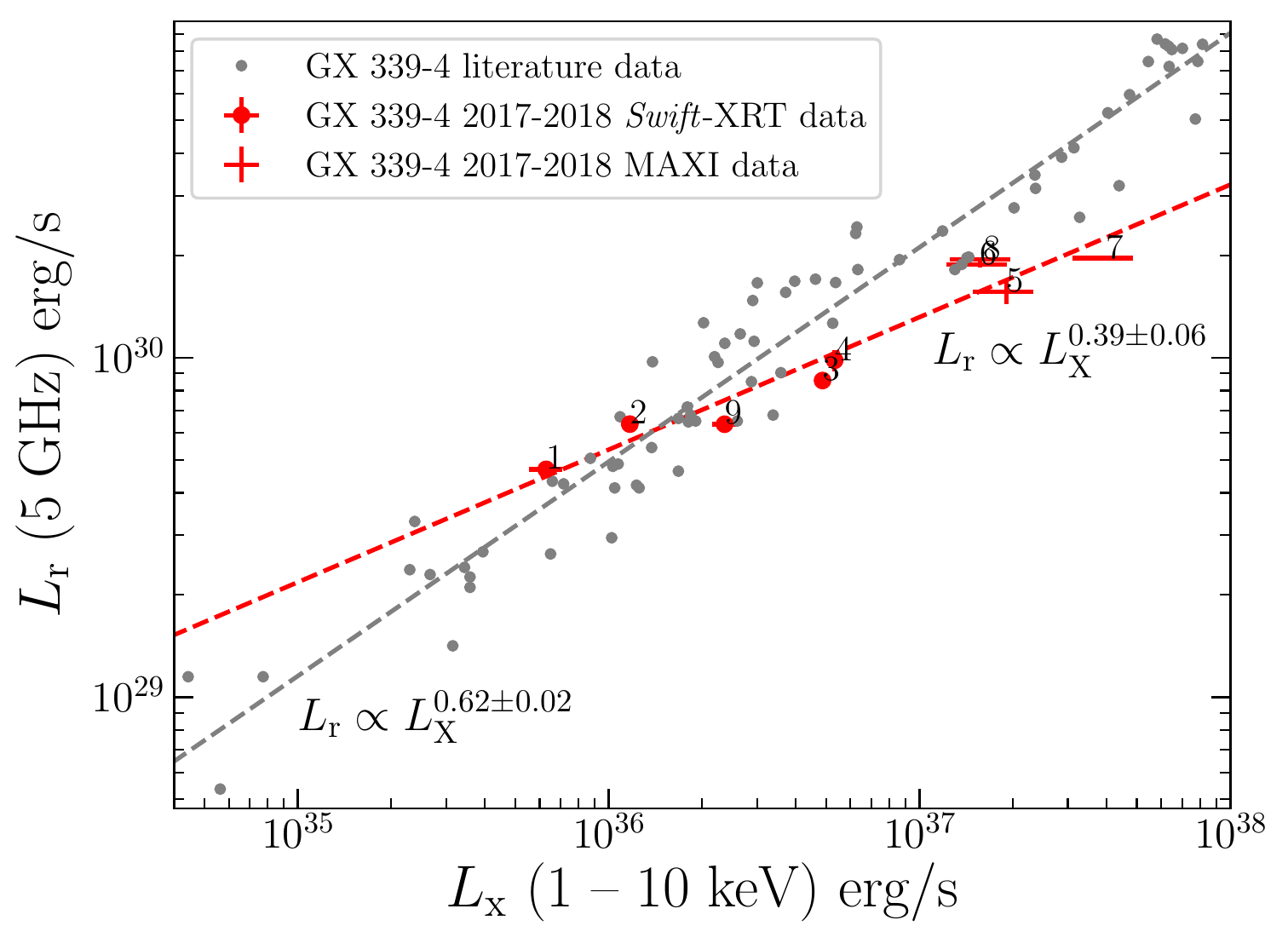}
\caption{The radio luminosity ($L_{\rm r}$) vs X-ray luminosity ($L_{\rm X}$) plot of \source, where the 5\,GHz radio luminosity was inferred from the 5.5\,GHz radio flux density and the corresponding radio spectral index. The grey points consist of literature outburst data from \source\,\citep{corbel}, while the red points correspond to data from the 2017--2018 hard-only outburst. The numbers near the red points indicate the order in which the data was obtained, so number 1 corresponds to the first observation of the outburst and number 9 corresponds to the last. As in shown in the figure, for the 2017--2018 outburst, the best-fit relation is $L_{\rm r} \propto L_{\rm X}^{0.39 \pm 0.06}$, flatter than successful outbursts combined from \source\ (where $L_{\rm r} \propto L_{\rm X}^{0.62 \pm 0.02}$).}
\label{LrLxnew}
\end{figure}

\subsection{The radio -- X-ray correlation}
\label{sec:lrlx}

\begin{figure*}
\centering
\includegraphics[width=1.0\textwidth]{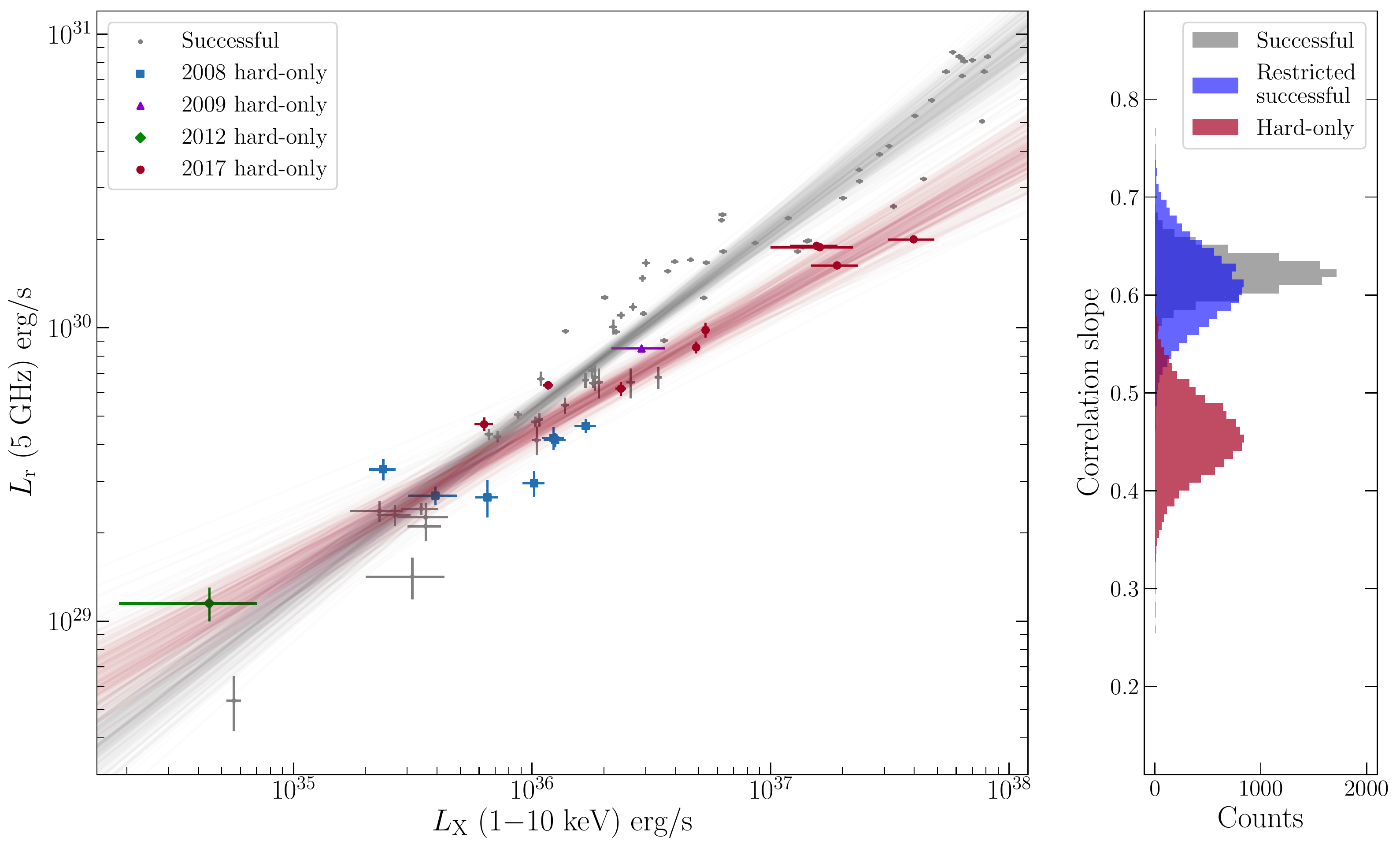}
\caption{\textit{(Left panel)} The radio and X-ray luminosities of \source, showing both hard-only outbursts (coloured squares, triangles, diamonds, and circles) and successful outburst data (grey dots). Data from the 2008, 2009 and 2012 hard-only outbursts, as well as the successful outburst luminosities were taken from \citet{corbel}. We determine a best-fit relation of $L_{\rm r} \propto L_{\rm X}^{0.46 \pm 0.04}$ (red fits), while the successful outburst data has a best-fit of $L_{\rm r} \propto L_{\rm X}^{0.62 \pm 0.02}$ (grey fits). \textit{(Right panel)} Histogram showing the fitting results for the successful (grey), restricted successful (blue), and hard-only (red) outburst correlation indexes. These results suggest that hard-only outbursts from \source\ show a flatter radio -- X-ray correlation than for successful outbursts.}
\label{LrLx2}
\end{figure*}

With our quasi-simultaneous radio and X-ray data (taken within 1\,day of each other)\footnote{The radio/X-ray luminosity data point from MJD~58058 was taken $>$1\,day apart (the mid-point of the X-ray observation was MJD 58058.1, while the ATCA radio observation was MJD~58059.9). Therefore, to best estimate the simultaneous radio luminosity, we fit the radio light curve assuming an exponential rise to estimate the 5.5\,GHz flux density at the time of the X-ray observation.}, we explored the correlation between the radio and X-ray emission. We used the \textsc{python} implementation of the linear regression algorithm \textsc{linmix\_err}\footnote{https://github.com/jmeyers314/linmix} \citep{2007ApJ...665.1489K} to perform a (10,000 iteration) linear fit to the radio and X-ray luminosities in logarithmic space, finding a flatter than usual radio -- X-ray coupling for \source, where $L_{\rm r} \propto L_{\rm X}^{0.39 \pm 0.06}$ for this hard-only outburst (Figure~\ref{LrLxnew}). This result included the \textit{MAXI} data which had an assumed value of $\Gamma$ (where $\Gamma$ = 1.5 -- 2.5). However, we note that due to the close proximity of the Sun, the MAXI images were poor. As such, excluding the \textit{MAXI} points did not change our results, providing a best-fit relation of $L_{\rm r} \propto L_{\rm X}^{0.3 \pm 0.1}$.

A flatter radio -- X-ray correlation may suggest a more inefficient coupling between the two emission mechanisms (where the dominant X-ray emission could now arise from synchrotron self-Compton emission from the base of the jet; e.g., \citealt{2005ApJ...635.1203M,2008MNRAS.389.1697C}; Shaw et al.\ 2020 submitted). An inefficient coupling may arise from a lower mass accretion rate, resulting in a lower magnetic flux and, therefore, possibly weaker jets \citep[e.g.,][]{1987PASJ...39..559S}. However, even though the observed relation does deviate significantly from that reported by \citet{corbel} and \citet{gallotracks}, such a deviation is not necessarily atypical. \citet{corbel} suggested that over luminosity ranges less than two orders of magnitude, the $L_{\rm r} - L_{\rm X}$ correlation index may deviate substantially (where indices of between 0 and 2 have been observed). Here, our radio and X-ray data only span a range of $\lesssim$1 order of magnitude in $L_{\rm r}$ and $\gtrsim$1.5 orders of magnitude in $L_{\rm X}$ (Figure~\ref{LrLxnew}).

To extend the X-ray luminosity range, we combined our 2017--2018 $L_{\rm r} - L_{\rm X}$ data with radio and X-ray luminosities taken from other hard-only outbursts from \source\ (Figure~\ref{LrLx2}, left panel), where we include data presented by \citealt{corbel} on the 2008, 2009 and 2012 hard-only outbursts. We find a best-fit relation of $L_{\rm r} \propto L_{\rm X}^{0.46 \pm 0.04}$ ($L_{\rm r} \propto L_{\rm X}^{0.44 \pm 0.07}$ without the \textit{MAXI} data), again much flatter than the $L_{\rm r} \propto L_{\rm X}^{0.62 \pm 0.02}$ relation for the successful outbursts from \source\ (similar to results from \citealt{corbel}, who analysed data from five successful and two hard-only outbursts). Performing a Monte-Carlo 2-dimensional 2-sample Kolmogorov-Smirnov (K-S) test\footnote{https://github.com/syrte/ndtest} \citep{1983MNRAS.202..615P,1987MNRAS.225..155F} on the successful and hard-only data sets yields a $p$-value of 0.031, providing statistical evidence that is suggestive that the hard-only and successful radio and X-ray luminosities do not arise from the same parent distribution. Confining the successful outburst data from \citet{corbel} so that the data are within our observed 2017--2018 radio and X-ray luminosities, we find a best fit of $L_{\rm r} \propto L_{\rm X}^{0.61 \pm 0.04}$ (Figure~\ref{LrLx2}, right panel), showing that there isn't a consistent deviation at these X-ray luminosities. 

Our findings indicate that $L_{\rm r} - L_{\rm X}$ correlations for hard-only outbursts may be flatter than the correlations exhibited by successful outbursts from \source, possibly acting as a indicator for outbursts remaining within the hard state. However, our radio and sub-mm monitoring indicate that the compact jet was variable within single observations (Section~\ref{sec:intra_obs_variability}). Therefore, without strictly simultaneous radio and X-ray observations, such variability may impact our findings. Additionally, \citet{2019ApJ...871...26K} find steeper correlation slopes when the X-ray band is extended from 3--9\,keV to 3--200\,keV, but as our comparisons were all made within the same band, our results should not be significantly affected. We note that because the X-ray observations need to span $>$2-orders of magnitude in luminosity, using this method as a precursor indicator may be impractical (as hard-only outbursts often do not extend to this full range). However, this method may allow some early notification (after $\sim$1-order of magnitude) that it could be a hard-only outburst, and could also be used to classify unknown historical outbursts, or outbursts with no X-ray spectral or timing observations.

When comparing our result to those from hard-only outbursts of other systems, unfortunately the picture is not very clear. To date, there is a dearth of radio -- X-ray monitoring campaigns of both successful and hard-only outbursts from a single source meaning that clear comparisons are difficult to make. One of the only other sources that has been well monitored during both types of outburst is the BH-LMXB H1743$-$322. However, this system displayed a much more complex radio -- X-ray connection, where it lies on the outlier track of the correlation. Sources on the outlier track exhibit a steeper correlation at higher X-ray luminosities, a flatter correlation index at lower X-ray luminosities, and seem to follow the standard correlation track below some critical X-ray luminosity \citep[e.g.,][]{2011MNRAS.414..677C,2019ApJ...871...26K}. As such, for hard-only outbursts (which remain at a relatively low X-ray luminosity) it showed a very shallow correlation (of $L_{\rm X}^{0.18 \pm 0.01}$; \citealt{2010MNRAS.401.1255J}), but for successful outbursts, the lower luminosity data follow a similarly shallow path ($L_{\rm X}^{0.23 \pm 0.07}$) before transitioning to a much steeper correlation ($L_{\rm X}^{1.38 \pm 0.03}$; \citealt{2011MNRAS.414..677C}). The transition has been attributed to switching between a radiatively efficient and inefficient accretion regime. \citet{2020MNRAS.491L..29W} presented radio and X-ray observations of H1743$-$322 during its 2018 hard-only outburst. These observations show that hard-only outbursts can transition between the two paths, following a very similar behaviour. These studies highlight that H1743$-$322 has appeared to follow the same radio -- X-ray correlation path regardless of the outburst being successful or not. Therefore, while \source\ may exhibit a deviance from its standard correlation during hard-only outbursts, such behaviour may not be as clear for the population of BH-LMXB systems, or even for \source\ at higher hard state X-ray luminosities. Therefore, additional monitoring of both types of outburst from the same source with strictly simultaneous radio and X-ray observations is required to identify if a flatter correlation index is indeed a feature of hard-only outbursts from all sources, or if it is something related only to \source\ (or only a subset of BH-LMXBs). In particular, dedicated and more complete monitoring during both the rise and decay are needed, and are now more achievable due to regular optical monitoring (XB-NEWS;  \citealt{2019AN....340..278R}) providing an early trigger for radio and X-ray monitoring.

\section{Conclusion}
\label{sec:conclusion}

Our detailed analysis of X-ray and radio observations taken during the 2017--2018 hard-only outburst of \source, implied that despite the outburst not entering a soft state, \source\ appeared to show a relatively standard X-ray spectral evolution. While we detected a softer X-ray spectrum at low X-ray luminosities early on in the outburst, using comparisons between both successful and hard-only outbursts from this source we find that this behaviour is normal for both types of outburst (successful and hard-only). We suggest that the changes to the X-ray spectrum during that time were possibly caused by X-ray reprocessing in an optically thick medium, likely the accretion flow, or were due to optical depth changes in an optically-thin accretion flow. 

Our radio and X-ray monitoring showed a flatter than typical radio -- X-ray correlation for this outburst, where $L_{\rm r} \propto L_{\rm X}^{0.39 \pm 0.06}$. Combining our 2017--2018 outburst data with those from other hard-only outbursts from this system also yielded a flatter than typical correlation, where $L_{\rm r} \propto L_{\rm X}^{0.46 \pm 0.04}$, which may arise from a more inefficient coupling between the accretion flow and the jets, possibly due to a lower mass accretion rate. A flatter radio -- X-ray correlation at these lower X-ray luminosities may indicate whether a source completes a full outburst cycle or remains only in the hard state, although extending the luminosity range beyond 2-orders of magnitude in the X-rays to ensure reliability means that this method may not be a useful precursor indicator, but it could be used to suggest a hard-only outburst as a high possibility early in the outburst (after $\sim$1 order of magnitude in X-ray luminosity), or for historical and new outbursts with limited X-ray monitoring such that the spectral state changes are not known. However, our radio and X-ray observations were not strictly simultaneous and we detected considerable intra-observational compact jet variability in the sub-mm band (and possibly some in the radio band). Additionally, our results explore a limited luminosity range, focusing on the low-luminosity hard state, which must be extended through better sampling. Also, this behaviour doesn't necessarily appear to be universal among other systems. Therefore, while a flatter correlation index may act as an indicator for a hard-only outburst from \source, this may not be the case for other systems. Hence, further monitoring is required to test its universality.

\section{Acknowledgements} 
We thank the anonymous referee for their helpful comments that improved this manuscript. We also thank the UvA Jetset group for useful discussions. TDR acknowledges support from the Netherlands Organisation for Scientific Research (NWO) Veni Fellowship, grant number 639.041.646, and financial contribution from the agreement ASI-INAF n.2017-14-H.0. SM is supported by an NWO Vici grant, grant number 639.043.513. ND, JvdE, and ASP are supported by a Vidi grant from NWO, awarded to ND. JvdE is supported by a Lee Hysan Junior Research Fellowship from St Hilda's College, Oxford. GRS acknowledges support from Natural Sciences and Engineering Research Council of Canada (NSERC) Discovery Grants (RGPIN-06569-2016).
The Australia Telescope Compact Array (ATCA) is part of the Australia Telescope National Facility which is funded by the Australian Government for operation as a National Facility managed by CSIRO. We acknowledge the Gomeroi people as the traditional owners of the ATCA Observatory site. This paper makes use of the following ALMA data: ADS/JAO.ALMA\#2017.1.00864.T. ALMA is a partnership of ESO (representing its member states), NSF (USA) and NINS (Japan), together with NRC (Canada), MOST and ASIAA (Taiwan), and KASI (Republic of Korea), in cooperation with the Republic of Chile. The Joint ALMA Observatory is operated by ESO, AUI/NRAO and NAOJ. The National Radio Astronomy Observatory is a facility of the National Science Foundation operated under cooperative agreement by Associated Universities, Inc. This research has made use of (i) NASA's Astrophysics Data System, (ii) data, software, and/or web tools obtained from the High Energy Astrophysics Science Archive Research Center (HEASARC), a service of the Astrophysics Science Division at NASA Goddard Space Flight Center (GSFC) and of the Smithsonian Astrophysical Observatory's High Energy Astrophysics Division, (iii) data supplied by the UK \textit{Swift} Science Data Centre at the University of Leicester, and (iiii) MAXI data provided by RIKEN, JAXA and the MAXI team.

\section*{Data Availability}
Data from \swift\ and \textit{MAXI} are publicly available from HEASARC (\url{https://heasarc.gsfc.nasa.gov/}), obsIDs are provided in the data tables. Results and best-fit parameters are all provided in this paper. Raw radio data can be accessed online (\url{https://atoa.atnf.csiro.au/query.jsp}), under project code C3057. Raw ALMA data are available online (\url{https://almascience.eso.org/asax/}) under project code 2017.1.00864.T. All radio and sub-mm results are provided in this paper.

\bibliographystyle{mnras}
\bibliography{bib}

\newpage
\onecolumn
\appendix
\label{sec:appendix}

\section{X-ray observation IDs and fitted parameters}
Fitted X-ray parameters and observation identification numbers (obsID) for all \Swift-XRT data from the 2013 hard-only outburst (Table~\ref{xraydatacomplete2013}) and 2018--2019 hard-only flare (Table~\ref{xraydatacomplete2018}).
\FloatBarrier

\begin{table}
  \begin{center}
    \caption{X-ray evolution of \source\ during its 2013 hard-only outburst. These parameters were determined from \Swift-XRT monitoring, where $N_{\rm H} = (0.56 \pm 0.01) \times 10^{22}$ cm$^{-2}$. The total $\chi^2$ for the joint fit is 8602.36 with 8619 degrees of freedom. Errors are one sigma.}
    \label{xraydatacomplete2013}
    \begin{tabular}{ccccccc}
    \hline
          Date & MJD & ObsID & Count rate & $\Gamma$ & Normalisation & Unabsorbed flux \\
          & & & (0.5 -- 10 keV) & & & (0.5 -- 10 keV) \\
          & & & & & & $\times 10^{-10}$ erg s$^{-1}$ cm$^{-2}$ \\
      \hline
      2013--08--06 & 56510 & 00032490012 & 1.21 $\pm$ 0.04 & 1.42 $\pm$ 0.06 & 0.009 $\pm$ 0.001 & 0.81 $\pm$ 0.03 \\
        2013--08--07 & 56511 & 00032898001 & 2.84 $\pm$ 0.06 & 1.31 $\pm$ 0.03 & 0.021 $\pm$ 0.001 & 2.05 $\pm$ 0.05 \\
      2013--08--08 & 56512 & 00032490013 & 4.07 $\pm$ 0.05 & 1.39 $\pm$ 0.02 & 0.031 $\pm$ 0.001 & 2.78 $\pm$ 0.04 \\
      2013--08--09 & 56513 & 00032898002 & 4.53 $\pm$ 0.07 & 1.37 $\pm$ 0.03 & 0.033 $\pm$ 0.001 & 3.06 $\pm$ 0.06 \\
      2013--08--10 & 56514 & 00032490014 & 5.77 $\pm$ 0.08 & 1.36 $\pm$ 0.03 & 0.044 $\pm$ 0.001 & 4.05 $\pm$ 0.07 \\
      2013--08--12 & 56516 & 00032490015 & 4.63 $\pm$ 0.07 & 1.38 $\pm$ 0.03 & 0.060 $\pm$ 0.002 & 5.44 $\pm$ 0.10 \\
      2013--08--15 & 56519 & 00032898005 & 9.60 $\pm$ 0.09 &  1.42 $\pm$ 0.02 & 0.075 $\pm$ 0.001 & 6.49 $\pm$ 0.08 \\
      2013--08--19 & 56523 & 00032898006 & 13.90 $\pm$ 0.12 & 1.50 $\pm$ 0.02 & 0.117 $\pm$ 0.002 & 9.22 $\pm$ 0.10 \\
      2013--08--21 & 56525 & 00032898007 & 10.54 $\pm$ 0.11 & 1.50 $\pm$ 0.02 & 0.172 $\pm$ 0.003 & 13.51 $\pm$ 0.17 \\
      2013--08--24 & 56528 & 00080180002 & 17.39 $\pm$ 0.16 & 1.52 $\pm$ 0.02 & 0.149 $\pm$ 0.003 & 11.55 $\pm$ 0.13 \\
      2013--08--27 & 56531 & 00032898010 & 23.55 $\pm$ 0.15 & 1.55 $\pm$ 0.01 & 0.215 $\pm$ 0.003 & 15.95 $\pm$ 0.12 \\
      2013--09--02 & 56537 & 00032898013 & 24.49 $\pm$ 0.15 & 1.54 $\pm$ 0.01 & 0.226 $\pm$ 0.003 & 17.00 $\pm$ 0.12 \\
      2013--09--08 & 56543 & 00032898016 & 31.42 $\pm$ 0.20 & 1.55 $\pm$ 0.01 & 0.279 $\pm$ 0.004 & 20.75 $\pm$ 0.15 \\
      2013--09--14 & 56549 & 00032898019 & 27.09 $\pm$ 0.18 & 1.54 $\pm$ 0.01 & 0.240 $\pm$ 0.003 & 18.12 $\pm$ 0.14 \\
      2013--09--22 & 56557 & 00032898023 & 26.27 $\pm$ 0.15 & 1.55 $\pm$ 0.01 & 0.238 $\pm$ 0.003 & 17.75 $\pm$ 0.12 \\
      2013--09--28 & 56563 & 00032898026 & 20.61 $\pm$ 0.14 & 1.54 $\pm$ 0.01 & 0.186 $\pm$  0.003 & 13.95 $\pm$ 0.12 \\
      2013--10--06 & 56571 & 00032898030 & 12.74 $\pm$ 0.11 & 1.55 $\pm$ 0.02 & 0.118 $\pm$ 0.002 & 8.77 $\pm$ 0.09 \\
      2013--10--12 & 56577 & 00032898033 & 7.56 $\pm$ 0.15 & 1.45 $\pm$ 0.04 & 0.065 $\pm$ 0.002 & 5.41 $\pm$ 0.13 \\
      2013--10--18 & 56583 & 00032898036 & 3.34 $\pm$ 0.06 & 1.45 $\pm$ 0.04 & 0.027 $\pm$ 0.001 & 2.29 $\pm$ 0.05 \\

      \hline
    \end{tabular}
  \end{center}

\end{table}

\begin{table}
  \begin{center}
    \caption{X-ray evolution of \source\ during its 2018--2019 hard-only flare. These parameters were determined from \Swift-XRT monitoring, where $N_{\rm H} = (0.56 \pm 0.01) \times 10^{22}$ cm$^{-2}$. The total $\chi^2$ for the joint fit is 3952.25 with 4163 degrees of freedom, the reduced $\chi^2$ for the joint fit is then $\simeq 0.95$. Errors are one sigma.}
    \label{xraydatacomplete2018}
    \begin{tabular}{ccccccc}
    \hline
          Date & MJD & ObsID & Count rate & $\Gamma$ & Normalisation & Unabsorbed flux \\
          & & & (0.5 -- 10 keV) & & & (0.5 -- 10 keV) \\
          & & & & & & $\times 10^{-10}$ erg s$^{-1}$ cm$^{-2}$ \\
      \hline
      2018--11--01 & 58423 & 00032898181 & 0.028 $\pm$ 0.004 & 1.65 $\pm$ 0.44 & 0.0004 $\pm$ 0.0001 & 0.024 $\pm$ 0.005 \\
      2019--01--21 & 58504 & 00032898182 & 11.19 $\pm$ 0.11 & 1.45 $\pm$ 0.02 & 0.089 $\pm$ 0.002 & 7.39 $\pm$ 0.09 \\    
      2019--01--22 & 58505 & 00032898183 & 10.21 $\pm$ 0.17 & 1.54 $\pm$ 0.03 & 0.085 $\pm$ 0.003 & 6.38 $\pm$ 0.13 \\
      2019--01--24 & 58507 & 00032898184 & 10.88 $\pm$ 0.14 & 1.42 $\pm$ 0.02 & 0.084 $\pm$ 0.002 & 7.23 $\pm$ 0.11 \\
      2019--01--27 & 58510 & 00032898185 & 10.39 $\pm$ 0.11 & 1.45 $\pm$ 0.02 & 0.083 $\pm$ 0.002 & 6.93 $\pm$ 0.09 \\
      2019--01--29 & 58512 & 00032898186 & 10.27 $\pm$ 0.10 & 1.47 $\pm$ 0.02 & 0.096 $\pm$ 0.002 & 7.83 $\pm$ 0.09 \\
      2019--02-02 & 58516 & 00032898188 & 4.67 $\pm$ 0.11 & 1.40 $\pm$ 0.04 & 0.091 $\pm$ 0.004 & 8.03 $\pm$ 0.22 \\
      2019--02--05 & 58519 & 00032898189 & 10.60 $\pm$ 0.10 & 1.49 $\pm$ 0.02 & 0.085 $\pm$ 0.002 & 6.79 $\pm$ 0.08 \\
      2019--02--12 & 58526 & 00032898190 & 7.47 $\pm$ 0.16 & 1.46 $\pm$ 0.04 & 0.071 $\pm$ 0.003 & 5.88 $\pm$ 0.16 \\
      2019--02--19 & 58533 & 00032898191 & 8.37 $\pm$ 0.12 & 1.44 $\pm$ 0.03 & 0.065 $\pm$ 0.002 & 5.47 $\pm$ 0.09 \\
      2019--02--26 & 58540 & 00032898194 & 6.65 $\pm$ 0.20 & 1.46 $\pm$ 0.06 & 0.052 $\pm$ 0.003 & 4.25 $\pm$ 0.15 \\
      2019--03--01 & 58543 & 00032898195 & 6.34 $\pm$ 0.23 & 1.31 $\pm$ 0.07 & 0.044 $\pm$ 0.003 & 4.36 $\pm$ 0.20 \\
      2019--03--10 & 58552 & 00032898199 & 5.98 $\pm$ 0.23 & 1.47 $\pm$ 0.08 & 0.046 $\pm$ 0.003 & 3.75 $\pm$ 0.18 \\
      2019--03--16 & 58558 & 00032898202 & 0.83 $\pm$ 0.02 & 1.24 $\pm$ 0.05 & 0.038 $\pm$ 0.002 & 4.16 $\pm$ 0.15 \\
      2019--03--27 & 58569 & 00032898205 & 0.52 $\pm$ 0.02 & 1.29 $\pm$ 0.06 & 0.033 $\pm$ 0.002 & 3.31 $\pm$ 0.13 \\
      2019--04--02 & 58575 & 00032898206 & 0.61 $\pm$ 0.02 & 1.22 $\pm$ 0.06 & 0.021 $\pm$ 0.001 & 2.30 $\pm$ 0.10 \\
      2019--04--09 & 58582 & 00032898207 & 0.68 $\pm$ 0.02 & 1.32 $\pm$ 0.05 & 0.023 $\pm$ 0.001 & 2.20 $\pm$ 0.08 \\
      2019--04--16 & 58589 & 00032898208 & 0.51 $\pm$ 0.02 & 1.26 $\pm$ 0.08 & 0.014 $\pm$ 0.001 & 1.52 $\pm$ 0.08 \\
      2019--04--24 & 58597 & 00032898209 & 0.39 $\pm$ 0.02 & 1.46 $\pm$ 0.11 & 0.013 $\pm$ 0.001 & 1.10 $\pm$ 0.07 \\
      2019--05--08 & 58611 & 00032898211 & 0.52 $\pm$ 0.03 & 1.43 $\pm$ 0.10 & 0.006 $\pm$ 0.001 & 0.55 $\pm$ 0.04 \\
      2019--05--14 & 58617 & 00032898212 & 0.40 $\pm$ 0.02 & 1.60 $\pm$ 0.11 & 0.005 $\pm$ 0.001 & 0.37 $\pm$ 0.02 \\

      \hline
    \end{tabular}
  \end{center}

\end{table}

\newpage

\section{Radio and sub-mm variability}
\label{sec:appendix_variability}

\begin{figure*}
\begin{center}
  \includegraphics[width=1\textwidth]{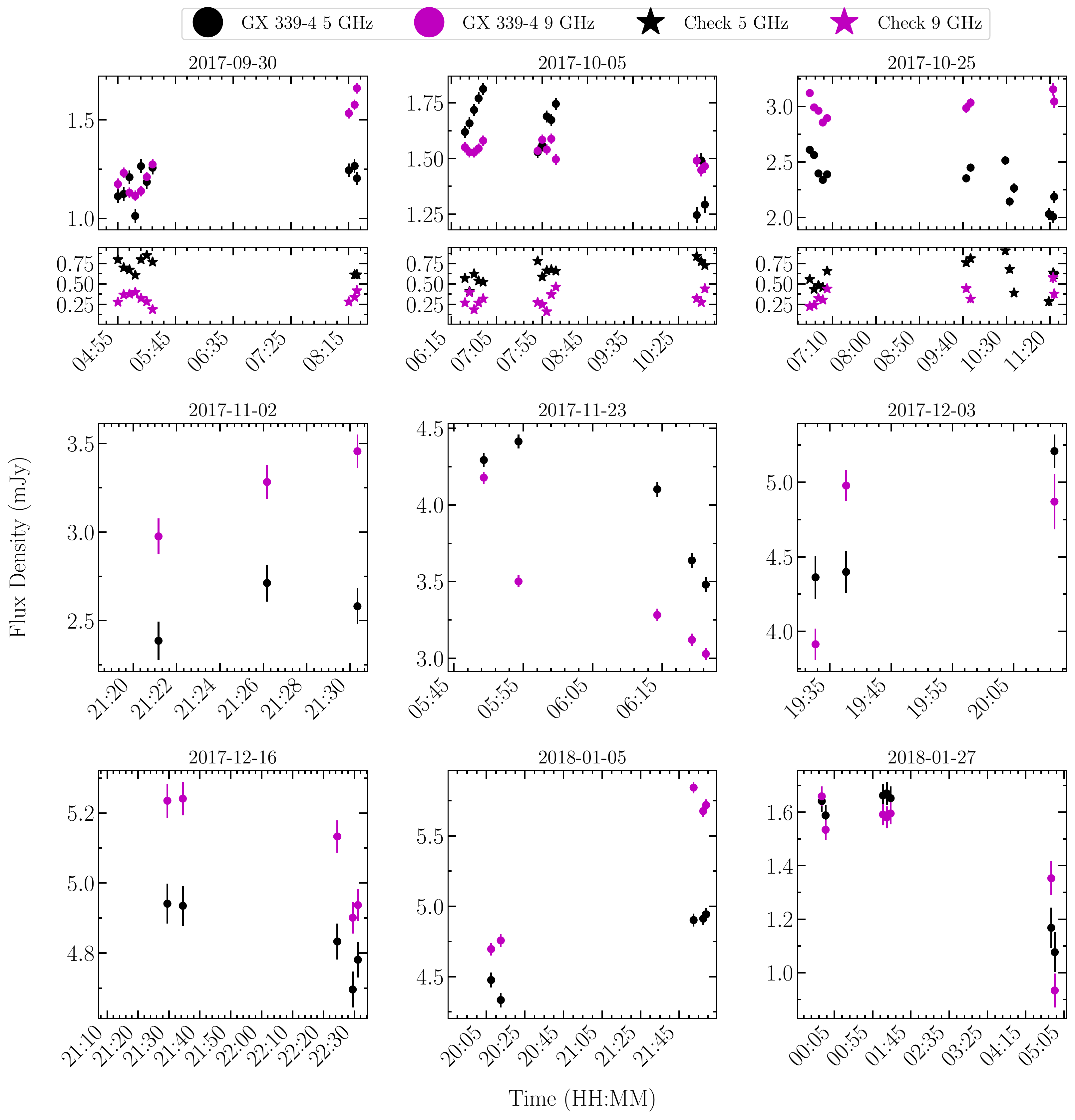}\\
 \caption{\label{fig:radiolc_var} Time-resolved radio lightcurves of GX 339--4 observed with ATCA at 5 and 9 GHz (black and magenta circles). We also show lightcurves of a check source in the ATCA field of view for three epochs when this source was detected (black and magenta stars). The radio emission we observe from GX 339--4 shows some short-timescale variability, in the form of increasing/decreasing trends within (e.g., 2017-10-05) and throughout some observations (e.g., 2017-09-30 and 2017-11-23), but no clear flaring activity. However, we caution that our check source can at times be just as variable as GX 339--4, which suggests the variations we observe in GX 339-4 may not be completely intrinsic to the source.}
\end{center}
 \end{figure*}

\begin{figure*}
\begin{center}
  \includegraphics[width=1\textwidth]{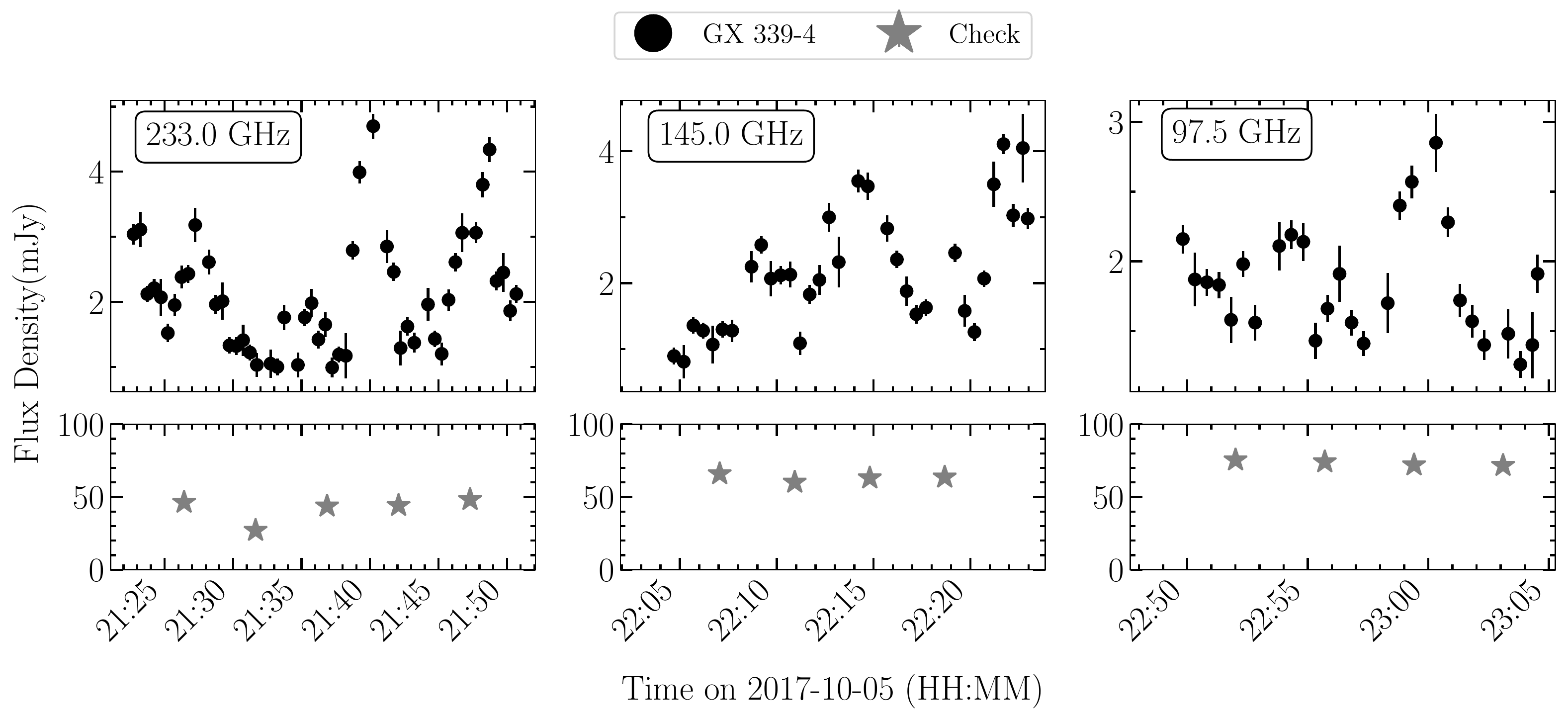}\\
 \caption{\label{fig:submmlc_var}Time-resolved sub-mm lightcurves of GX~339$-$4 observed with ALMA at 97.5, 145, and 233 GHz ({\it top panels}, black circles). We also show lightcurves of the check source, J1631--5256 ({\it bottom panels}, each gray star is a single $\sim20$ sec scan on the check source). The sub-mm emission we observe from GX~339$-$4 is clearly variable, where we observe multiple, structured flaring events in all the sampled bands. The check source remains relatively constant throughout the observations (any variations are $<10$ \% of the average flux density of the check source), indicating that the variability we observe in GX~339$-$4 is likely intrinsic to the source, and not due to atmospheric or instrumental effects.}
\end{center}
 \end{figure*}

Intra-observation lightcurves for the ATCA radio (Table~\ref{tab:radio_intra_obs_data} and Figure~\ref{fig:radiolc_var}) and ALMA sub-mm observations (Table~\ref{tab:submm_intra_obs_data} and Figure~\ref{fig:submmlc_var}).

\begin{center}
\begin{longtable}{clccc}
\caption {Time-resolved ATCA radio data of \source\ for each of the observational epochs. The MJD represents the mid-point of the 5-minute snapshot.} 
\label{tab:radio_intra_obs_data} \\

\hline 
  \multicolumn{1}{c}{Epoch}& \multicolumn{1}{c}{MJD} & \multicolumn{1}{c}{Central frequency}   & \multicolumn{1}{c}{$S_{\nu}$} & \multicolumn{1}{c}{Error}  \\

 \multicolumn{1}{c}{} & \multicolumn{1}{c}{} & \multicolumn{1}{c}{(GHz)} & \multicolumn{1}{c}{(mJy)} & \multicolumn{1}{c}{(mJy)} \\
 
 \hline

\endfirsthead

\multicolumn{5}{c}%
{{\tablename\ \thetable{} -- Continued from previous page. Intra-observational ATCA flux densities.}} \\

\hline 

  \multicolumn{1}{c}{Epoch}& \multicolumn{1}{c}{MJD} & \multicolumn{1}{c}{Central frequency}   & \multicolumn{1}{c}{$S_{\nu}$} & \multicolumn{1}{c}{Error}  \\

 \multicolumn{1}{c}{} & \multicolumn{1}{c}{} & \multicolumn{1}{c}{(GHz)} & \multicolumn{1}{c}{(mJy)} & \multicolumn{1}{c}{(mJy)} \\
 
 \hline

\endhead
\hline
\multicolumn{5}{c}{{Continued on next page}} \\ \hline
\endfoot

\hline
\endlastfoot

2017-09-30 & 58026.20496527777 & 5.5 & 1.112 & 0.034 \\
& 58026.20843750001 & 5.5 & 1.124 & 0.035 \\
& 58026.211909722224 & 5.5 & 1.209 & 0.035 \\
& 58026.21538194444 & 5.5 & 1.012 & 0.035 \\
& 58026.21885416666 & 5.5 & 1.265 & 0.035 \\
& 58026.222326388895 & 5.5 & 1.185 & 0.034 \\
& 58026.22579861112 & 5.5 & 1.257 & 0.034 \\
& 58026.34385416666 & 5.5 & 1.244 & 0.034 \\
& 58026.347326388895 & 5.5 & 1.266 & 0.034 \\
& 58026.348715277774 & 5.5 & 1.203 & 0.034 \\
& 58026.20496527777 & 9.0 & 1.174 & 0.028 \\
& 58026.20843750001 & 9.0 & 1.231 & 0.027 \\
& 58026.211909722224 & 9.0 & 1.130 & 0.027 \\
& 58026.21538194444 & 9.0 & 1.114 & 0.027 \\
& 58026.21885416666 & 9.0 & 1.139 & 0.027 \\
& 58026.222326388895 & 9.0 & 1.21 & 0.027 \\
& 58026.22579861112 & 9.0 & 1.274 & 0.027 \\
& 58026.34385416666 & 9.0 & 1.534 & 0.027 \\
& 58026.347326388895 & 9.0 & 1.577 & 0.027 \\
& 58026.348715277774 & 9.0 & 1.661 & 0.027 \\

\hline

2017-10-05 & 58031.27093750001 & 5.5 & 1.619 & 0.026 \\
& 58031.274409722224 & 5.5 & 1.66 & 0.027 \\
& 58031.27788194444 & 5.5 & 1.720 & 0.027 \\
& 58031.28135416666 & 5.5 & 1.770 & 0.027 \\
& 58031.28482638889 & 5.5 & 1.812 & 0.027 \\
& 58031.32649305556 & 5.5 & 1.528 & 0.027 \\
& 58031.32996527777 & 5.5 & 1.560 & 0.027 \\
& 58031.33343750001 & 5.5 & 1.689 & 0.027 \\
& 58031.336909722224 & 5.5 & 1.673 & 0.027 \\
& 58031.34038194444 & 5.5 & 1.745 & 0.027 \\
& 58031.44802083333 & 5.5 & 1.246 & 0.035 \\
& 58031.45149305556 & 5.5 & 1.491 & 0.035 \\
& 58031.45427083333 & 5.5 & 1.293 & 0.036 \\
& 58031.27093750001 & 9.0 & 1.550 & 0.023 \\
& 58031.274409722224 & 9.0 & 1.528 & 0.023 \\
& 58031.27788194444 & 9.0 & 1.527 & 0.023 \\
& 58031.28135416666 & 9.0 & 1.545 & 0.023 \\
& 58031.28482638889 & 9.0 & 1.580 & 0.023 \\
& 58031.32649305556 & 9.0 & 1.534 & 0.024 \\
& 58031.32996527777 & 9.0 & 1.584 & 0.024 \\
& 58031.33343750001 & 9.0 & 1.540 & 0.024 \\
& 58031.336909722224 & 9.0 & 1.588 & 0.024 \\
& 58031.34038194444 & 9.0 & 1.496 & 0.024 \\
& 58031.44802083333 & 9.0 & 1.490 & 0.028 \\
& 58031.45149305556 & 9.0 & 1.450 & 0.028 \\
& 58031.45427083333 & 9.0 & 1.465 & 0.029 \\

\hline

2017-10-25 & 58051.280428240745 & 5.5 & 2.61 & 0.028 \\
& 58051.28390046296 & 5.5 & 2.563 & 0.028 \\
& 58051.2873726852 & 5.5 & 2.397 & 0.028 \\
& 58051.29084490741 & 5.5 & 2.339 & 0.028 \\
& 58051.29431712962 & 5.5 & 2.390 & 0.028 \\
& 58051.405428240745 & 5.5 & 2.353 & 0.038 \\
& 58051.40890046296 & 5.5 & 2.449 & 0.039 \\
& 58051.43667824073 & 5.5 & 2.514 & 0.042 \\
& 58051.44015046296 & 5.5 & 2.144 & 0.043 \\
& 58051.44362268518 & 5.5 & 2.264 & 0.043 \\
& 58051.47140046296 & 5.5 & 2.031 & 0.053 \\
& 58051.4748726852 & 5.5 & 2.007 & 0.055 \\
& 58051.47568287038 & 5.5 & 2.187 & 0.055 \\
& 58051.280428240745 & 9.0 & 3.121 & 0.032 \\
& 58051.28390046296 & 9.0 & 2.992 & 0.032 \\
& 58051.2873726852 & 9.0 & 2.962 & 0.031 \\
& 58051.29084490741 & 9.0 & 2.856 & 0.032 \\
& 58051.29431712962 & 9.0 & 2.895 & 0.031 \\
& 58051.405428240745 & 9.0 & 2.986 & 0.044 \\
& 58051.40890046296 & 9.0 & 3.034 & 0.044 \\
& 58051.4748726852 & 9.0 & 3.155 & 0.057 \\
& 58051.47568287038 & 9.0 & 3.045 & 0.058 \\

\hline

2017-11-02 & 58059.889918981484 & 5.5 & 2.386 & 0.109 \\
& 58059.89339120369 & 5.5 & 2.712 & 0.104 \\
& 58059.89628472223 & 5.5 & 2.581 & 0.101 \\
& 58059.889918981484 & 9.0 & 2.976 & 0.101 \\
& 58059.89339120369 & 9.0 & 3.283 & 0.096 \\
& 58059.89628472223 & 9.0 & 3.457 & 0.094 \\

\hline
2017-11-23 & 58080.242581018516 & 5.5 & 4.293 & 0.045 \\
& 58080.246053240735 & 5.5 & 4.414 & 0.045 \\
& 58080.259942129625 & 5.5 & 4.102 & 0.048 \\
& 58080.263414351844 & 5.5 & 3.638 & 0.048 \\
& 58080.26480324073 & 5.5 & 3.48 & 0.047 \\
& 58080.242581018516 & 9.0 & 4.178 & 0.039 \\
& 58080.246053240735 & 9.0 & 3.501 & 0.039 \\
& 58080.259942129625 & 9.0 & 3.282 & 0.041 \\
& 58080.263414351844 & 9.0 & 3.120 & 0.040 \\
& 58080.26480324073 & 9.0 & 3.028 & 0.040 \\

\hline

2017-12-03 & 58090.814340277764 & 5.5 & 4.363 & 0.145 \\
& 58090.81781249999 & 5.5 & 4.399 & 0.141 \\
& 58090.84142361111 & 5.5 & 5.209 & 0.111 \\
& 58090.814340277764 & 9.0 & 3.914 & 0.106 \\
& 58090.81781249999 & 9.0 & 4.978 & 0.104 \\
& 58090.84142361111 & 9.0 & 4.870 & 0.186 \\

\hline
2017-12-16 & 58103.89547453704 & 5.5 & 4.941 & 0.057 \\
& 58103.898946759255 & 5.5 & 4.935 & 0.057 \\
& 58103.93366898148 & 5.5 & 4.833 & 0.051 \\
& 58103.93714120369 & 5.5 & 4.696 & 0.051 \\
& 58103.938298611116 & 5.5 & 4.781 & 0.051 \\
& 58103.89547453704 & 9.0 & 5.235 & 0.048 \\
& 58103.898946759255 & 9.0 & 5.241 & 0.048 \\
& 58103.93366898148 & 9.0 & 5.133 & 0.046 \\
& 58103.93714120369 & 9.0 & 4.901 & 0.045 \\
& 58103.938298611116 & 9.0 & 4.937 & 0.045 \\

\hline
2018-01-05 & 58123.838530092595 & 5.5 & 4.476 & 0.053 \\
& 58123.842002314814 & 5.5 & 4.334 & 0.052 \\
& 58123.91144675926 & 5.5 & 4.902 & 0.046 \\
& 58123.914918981485 & 5.5 & 4.912 & 0.045 \\
& 58123.915960648155 & 5.5 & 4.943 & 0.045 \\
& 58123.838530092595 & 9.0 & 4.696 & 0.045 \\
& 58123.842002314814 & 9.0 & 4.757 & 0.045 \\
& 58123.91144675926 & 9.0 & 5.843 & 0.041 \\
& 58123.914918981485 & 9.0 & 5.676 & 0.040 \\
& 58123.915960648155 & 9.0 & 5.719 & 0.040 \\

\hline

2018-01-27 & 58145.992002314815 & 5.5 & 1.641 & 0.04 \\
& 58145.99547453704 & 5.5 & 1.588 & 0.04 \\
& 58146.04755787037 & 5.5 & 1.662 & 0.043 \\
& 58146.05103009259 & 5.5 & 1.670 & 0.043 \\
& 58146.05450231481 & 5.5 & 1.652 & 0.044 \\
& 58146.200335648144 & 5.5 & 1.168 & 0.075 \\
& 58146.203460648154 & 5.5 & 1.077 & 0.075 \\
& 58145.992002314815 & 9.0 & 1.659 & 0.037 \\
& 58145.99547453704 & 9.0 & 1.534 & 0.038 \\
& 58146.04755787037 & 9.0 & 1.591 & 0.040 \\
& 58146.05103009259 & 9.0 & 1.580 & 0.041 \\
& 58146.05450231481 & 9.0 & 1.595 & 0.041 \\
& 58146.200335648144 & 9.0 & 1.353 & 0.063 \\
& 58146.203460648154 & 9.0 & 0.934 & 0.063 \\

\hline

\end{longtable}

\end{center}

 \begin{center}
\begin{longtable}{lccc}
\caption {Time-resolved sub-mm lightcurve data of GX 339--4 observed with ALMA at 97.5, 145, and 233 GHz on 2017-10-05. The MJD represents the mid-point of the 30-second snapshot.} 
\label{tab:submm_intra_obs_data} \\

\hline 
  \multicolumn{1}{c}{MJD} & \multicolumn{1}{c}{Central frequency}   & \multicolumn{1}{c}{$S_{\nu}$} & \multicolumn{1}{c}{Error}  \\

 \multicolumn{1}{c}{} & \multicolumn{1}{c}{(GHz)} & \multicolumn{1}{c}{(mJy)} & \multicolumn{1}{c}{(mJy)} \\
 \hline

\endfirsthead

\multicolumn{4}{c}%
{{\tablename\ \thetable{} -- Continued from previous page. Intra-observational ALMA densities taken on 2017 October 05.}} \\

\hline 

  \multicolumn{1}{c}{MJD} & \multicolumn{1}{c}{Central frequency}   & \multicolumn{1}{c}{$S_{\nu}$} & \multicolumn{1}{c}{Error}  \\

  \multicolumn{1}{c}{} & \multicolumn{1}{c}{(GHz)} & \multicolumn{1}{c}{(mJy)} & \multicolumn{1}{c}{(mJy)} \\
 \hline

\endhead
\hline
\multicolumn{4}{c}{{Continued on next page}} \\ \hline
\endfoot

\hline
\endlastfoot

58031.891377314816 & 233.0 & 3.04 & 0.16\\
58031.89172453704  & 233.0 & 3.11 & 0.27\\
58031.89207175926  & 233.0 & 2.12 & 0.12\\
58031.89241898148  & 233.0 & 2.21 & 0.14\\
58031.8927662037   & 233.0 & 2.07 & 0.29\\
58031.893113425926 & 233.0 & 1.52 & 0.14\\
58031.89346064815  & 233.0 & 1.95 & 0.17\\
58031.89380787037  & 233.0 & 2.38 & 0.18\\
58031.894155092596 & 233.0 & 2.43 & 0.14\\
58031.89450231481  & 233.0 & 3.18 & 0.26\\
58031.89519675926  & 233.0 & 2.61 & 0.19\\
58031.89554398148  & 233.0 & 1.96 & 0.14\\
58031.895891203705 & 233.0 & 2.01 & 0.29\\
58031.89623842593  & 233.0 & 1.33 & 0.12\\
58031.896585648145 & 233.0 & 1.32 & 0.14\\
58031.89693287037  & 233.0 & 1.41 & 0.24\\
58031.89728009259  & 233.0 & 1.22 & 0.12\\
58031.897627314815 & 233.0 & 1.03 & 0.18\\
58031.89832175926  & 233.0 & 1.05 & 0.22\\
58031.89866898148  & 233.0 & 1.00 & 0.13\\
58031.8990162037   & 233.0 & 1.76 & 0.19\\
58031.89971064815  & 233.0 & 1.03 & 0.19\\
58031.90005787037  & 233.0 & 1.76 & 0.14\\
58031.900405092594 & 233.0 & 1.98 & 0.22\\
58031.90075231481  & 233.0 & 1.42 & 0.14\\
58031.90109953703  & 233.0 & 1.65 & 0.19\\
58031.90144675926  & 233.0 & 0.99 & 0.15\\
58031.90179398148  & 233.0 & 1.19 & 0.12\\
58031.902141203704 & 233.0 & 1.17 & 0.35\\
58031.90248842593  & 233.0 & 2.79 & 0.14\\
58031.90283564815  & 233.0 & 3.99 & 0.17\\
58031.90353009259  & 233.0 & 4.70 & 0.19\\
58031.90422453704  & 233.0 & 2.85 & 0.25\\
58031.90457175926  & 233.0 & 2.46 & 0.14\\
58031.90491898148  & 233.0 & 1.29 & 0.27\\
58031.90526620371  & 233.0 & 1.62 & 0.14\\
58031.90561342592  & 233.0 & 1.37 & 0.15\\
58031.90630787037  & 233.0 & 1.96 & 0.25\\
58031.90665509259  & 233.0 & 1.43 & 0.13\\
58031.907002314816 & 233.0 & 1.20 & 0.18\\
58031.90734953704  & 233.0 & 2.03 & 0.16\\
58031.907696759255 & 233.0 & 2.61 & 0.14\\
58031.90804398148  & 233.0 & 3.06 & 0.30\\
58031.908738425926 & 233.0 & 3.06 & 0.16\\
58031.90908564815  & 233.0 & 3.80 & 0.19\\
58031.90943287037  & 233.0 & 4.34 & 0.19\\
58031.90978009259  & 233.0 & 2.32 & 0.15\\
58031.91012731481  & 233.0 & 2.45 & 0.30\\
58031.910474537035 & 233.0 & 1.86 & 0.16\\
58031.910787037035 & 233.0 & 2.12 & 0.14\\

58031.9200462963   & 145.0 & 0.90 & 0.13\\
58031.920393518514 & 145.0 & 0.81 & 0.25\\
58031.92074074074  & 145.0 & 1.36 & 0.13\\
58031.92108796296  & 145.0 & 1.28 & 0.12\\
58031.921435185184 & 145.0 & 1.07 & 0.29\\
58031.92178240741  & 145.0 & 1.30 & 0.12\\
58031.92212962963  & 145.0 & 1.28 & 0.17\\
58031.92282407408  & 145.0 & 2.25 & 0.24\\
58031.923171296294 & 145.0 & 2.58 & 0.13\\
58031.92351851852  & 145.0 & 2.07 & 0.27\\
58031.92386574074  & 145.0 & 2.12 & 0.14\\
58031.924212962964 & 145.0 & 2.13 & 0.20\\
58031.92456018519  & 145.0 & 1.09 & 0.18\\
58031.92490740741  & 145.0 & 1.83 & 0.14\\
58031.92525462963  & 145.0 & 2.05 & 0.22\\
58031.92560185185  & 145.0 & 3.00 & 0.22\\
58031.92594907407  & 145.0 & 2.32 & 0.39\\
58031.92664351852  & 145.0 & 3.55 & 0.17\\
58031.92699074074  & 145.0 & 3.47 & 0.21\\
58031.92768518518  & 145.0 & 2.83 & 0.20\\
58031.928032407406 & 145.0 & 2.36 & 0.13\\
58031.92837962963  & 145.0 & 1.88 & 0.22\\
58031.92872685185  & 145.0 & 1.53 & 0.14\\
58031.929074074076 & 145.0 & 1.63 & 0.13\\
58031.93011574074  & 145.0 & 2.46 & 0.14\\
58031.93046296296  & 145.0 & 1.58 & 0.24\\
58031.930810185186 & 145.0 & 1.26 & 0.14\\
58031.93115740741  & 145.0 & 2.07 & 0.12\\
58031.931504629625 & 145.0 & 3.50 & 0.34\\
58031.93185185185  & 145.0 & 4.11 & 0.15\\
58031.93219907407  & 145.0 & 3.03 & 0.17\\
58031.932546296295 & 145.0 & 4.05 & 0.52\\
58031.9327199074   & 145.0 & 2.98 & 0.17\\

58031.95130787037 & 97.5  & 2.16 & 0.10  \\
58031.951655092584& 97.5  & 1.87 & 0.19 \\
58031.95200231481 & 97.5  & 1.85 & 0.10\\
58031.95234953703 & 97.5  & 1.83 & 0.10\\
58031.952696759254& 97.5  & 1.58 & 0.17\\
58031.95304398148 & 97.5  & 1.98 & 0.09\\
58031.9533912037  & 97.5  & 1.56 & 0.13\\
58031.95408564815 & 97.5  & 2.11 & 0.18\\
58031.95443287036 & 97.5  & 2.19 & 0.10\\
58031.95478009259 & 97.5  & 2.14 & 0.14\\
58031.95512731481 & 97.5  & 1.43 & 0.13\\
58031.95547453703 & 97.5  & 1.66 & 0.10\\
58031.95582175926 & 97.5  & 1.91 & 0.20\\
58031.95616898148 & 97.5  & 1.56 & 0.09\\
58031.956516203696& 97.5  & 1.41 & 0.09\\
58031.95721064814 & 97.5  & 1.70 & 0.22\\
58031.957557870366& 97.5  & 2.40 & 0.10\\
58031.95790509259 & 97.5  & 2.57 & 0.12\\
58031.958599537036& 97.5  & 2.85 & 0.21\\
58031.95894675926 & 97.5  & 2.28 & 0.11\\
58031.95929398148 & 97.5  & 1.72 & 0.12\\
58031.959641203706& 97.5  & 1.57 & 0.12\\
58031.95998842592 & 97.5  & 1.40 & 0.11\\
58031.96068287037 & 97.5  & 1.48 & 0.18\\
58031.96103009259 & 97.5  & 1.26 & 0.10\\
58031.961377314816& 97.5  & 1.40 & 0.24\\
58031.961527777785& 97.5  & 1.91 & 0.14\\

\end{longtable}

\end{center}

\bsp	
\label{lastpage}
\end{document}